\documentclass[fleqn,usenatbib]{mnras}

\usepackage{newtxtext,newtxmath}

\usepackage[T1]{fontenc}

\DeclareRobustCommand{\VAN}[3]{#2}
\let\VANthebibliography\thebibliography
\def\thebibliography{\DeclareRobustCommand{\VAN}[3]{##3}\VANthebibliography}

\usepackage{graphicx}	
\usepackage{amsmath}	
\usepackage{xcolor}

\title[JWST galaxy morphology at $0.6 < z < 8.0$ ]{Beyond the Disc--Spheroid Dichotomy at $0.6<z<8.0$: Evidence for a Morphological Indeterminacy Regime from JWST}

\author[Kolesnikov et al. 2026]{
I. Kolesnikov,$^{1}$\thanks{E-mail: igor.s.kolesnikov@gmail.com}
R. R. de Carvalho,$^{1,2}$
J. H. Lee,$^{3}$
R. Flores-Freitas,$^{1}$
V. M. Sampaio,$^{4, 5}$
L. A. Silva-Lima,$^{6}$
\\
$^{1}$ NAT - Universidade Cidade de São Paulo, 01506-000, SP, Brazil\\
$^{2}$ Universidade de São Paulo, IAG, Rua o Matão 1226, Cidade Universitária, São Paulo 05508-900, Brazil\\
$^{3}$ Research Institute of Basic Sciences, Seoul National University, Seoul 08826, Republic of Korea\\
$^{4}$ Instituto de Física, Universidad Técnica Federico Santa María, Av. España 1680, Valparaíso, Chile\\
$^{5}$ Millennium Nucleus for Galaxies (MINGAL), Santiago, Chile\\
$^{6}$ Laboratório Nacional de Astrofísica (LNA), Rua dos Estados Unidos 154, Bairro das Nações, 37504-364, Itajubá, MG, Brazil
}

\date{Accepted XX. Received XX/XX/2026; in original form XX/XX/2026}

\pubyear{2026}

\begin{document}
\label{firstpage}
\pagerange{\pageref{firstpage}--\pageref{lastpage}}
\maketitle

\begin{abstract}

Galaxy morphology is a fundamental tracer of galaxy assembly, but its characterization becomes increasingly uncertain in the early Universe. We investigate morphological classifications across different methodologies over the redshift range $0.6 < z < 8.0$, using JWST observations of 15,567 galaxies with 6,099 synthetic galaxies from IllustrisTNG. To quantify morphology without relying on visual labels or prescribed light profiles, we use an unsupervised-supervised hybrid machine-learning approach based on the non-parametric MEGG system, which combines the second moment of the light distribution, Entropy, Gini, and Gradient Pattern Analysis. We find that parametric and non-parametric classifications remain broadly consistent at $z < 2.5$, but diverge systematically at higher redshift. We identify this epoch as the Morphological Indeterminacy Regime (MIR), in which classical disc and spheroidal categories no longer provide a unique description of the galaxy population. The fraction of mixed-interpretation galaxies increases with redshift, reaching $F_{\rm MIR} \sim 0.55$ at $z > 2.5$. Cross-domain and projection-invariance tests show that the classifications are stable and insensitive to data set and viewing angle. In the interval $2.5 < z < 8.0$, MIR galaxies occupy an intermediate stellar-mass regime ($\rm median; \log(M_{\star}/M_{\odot}) \sim 9.7$), between low-mass discs ($\rm median; \log(M_{\star}/M_{\odot}) \sim 9.5$) and massive spheroids ($\rm median; \log(M_{\star}/M_{\odot}) \sim 10.1$), while irregular systems have stellar-mass distributions similar to those of MIR galaxies. We interpret these results as evidence that morphological indeterminacy is a characteristic feature of galaxy assembly in the early Universe, reflecting a population of dynamically young systems that have not yet settled into the classical Hubble sequence.

\end{abstract}



\section{Introduction}

Galaxy morphology is one of the most visible outcomes of galaxy formation, but also one of the hardest to interpret. A morphological class is never just a label. It condenses the cumulative effect of angular momentum acquisition, gas accretion, star formation, dynamical instabilities, merging, and environmental influence. For that reason, the evolution of the fraction of discs and spheroids with redshift has long been treated as an observational route to the physical history of galaxy assembly. Within the standard picture of hierarchical structure formation \citep{White1978,Blumenthal1984,Davis1985,Mo1998}, this question is especially relevant because discs and spheroids are not expected to arise from identical evolutionary pathways, nor on the same timescales.

The difficulty is that morphology is inferred from images, and what is called a disc or a spheroid depends not only on the galaxy itself, but also on image depth, angular resolution, rest-frame wavelength, and, importantly, on the method used to describe the light distribution. Classical visual inspection works well for nearby galaxies, but its extension to high redshift rapidly becomes problematic. Quantitative systems were introduced precisely to address this limitation, either through parametric descriptions of the surface-brightness profile or through non-parametric measurements such as concentration, asymmetry, clumpiness, Gini, and $M_{20}$ \citep{Conselice2003,Lotz2004,Conselice2014}. These approaches represented a major step forward, but they did not remove the problem. They made it explicit: morphology classification depends on which aspect of the structure is being measured.

This issue became central with the Hubble Space Telescope (HST) deep surveys. By the time of CANDELS, it was already clear that the high-redshift Universe could not be described as a scaled version of the nearby one. Many galaxies at $z \gtrsim 2$ are compact, asymmetric, clumpy, or morphologically disturbed, including massive systems \citep{Kartaltepe2015,HuertasCompany2015}. At the same time, several studies found evidence that ordered structures were already in place at early epochs, including galaxies with low S\'ersic indices, extended light distributions, or visual appearances suggestive of discs \citep{Bruce2012,vanderWel2014}. The literature therefore moved toward a less comfortable but more realistic picture: regularity emerges early in at least part of the population, but the interpretation of that regularity is not method-independent.

James Webb Space Telescope (JWST) pushed this discussion into a new regime. For the first time, it became possible to examine the rest-frame optical structure of galaxies well into the epoch where earlier work relied mainly on rest-frame ultraviolet (UV) morphology. This matters because UV light overemphasizes young star-forming regions, whereas optical light traces the older stellar distribution more directly. In principle, JWST should make the structural classification of distant galaxies more robust. In practice, it also made the ambiguity more visible: many early galaxies do not fit comfortably into a clean disc--spheroid dichotomy. Recent JWST studies have reported a substantial diversity of structures at high redshift, including compact systems, irregular objects, clumpy galaxies, and sources whose light profiles are consistent with disc-like configurations \citep{Robertson2023}. The improved data therefore did not remove the ambiguity; in many cases, they made it easier to see.

Recent results on morphological fractions should be read against this background. \citet{Lee2024} analyzed a JWST sample spanning a broad redshift interval and found that disc galaxies dominate over spheroids even at high redshift. This result is important because it suggests that disc-like configurations may become common much earlier than usually assumed. \citet{kolesnikov_paper2}, using a hybrid unsupervised--supervised framework based on non-parametric morphology, and without visual labels entering the construction of the training sets, also examined the redshift evolution of morphological fractions. Although their analysis covers a lower-redshift interval, their results are consistent with \citet{Lee2024}. Other recent studies have also explored machine-defined morphological classifications without relying on a purely visual taxonomy. \citet{Dai2023} applied a two-step clustering scheme to HST/COSMOS-DASH galaxies, combining an unsupervised stage with a supervised classifier trained on the classes generated in the first step, while \citet{Tohill2024} used an unsupervised deep-learning approach to identify self-similar morphological groups in JWST galaxies out to $z\sim8$. These studies are relevant precedents for the broader move toward non-visual classification schemes, even if they do not all address the redshift evolution of disc and spheroid fractions in the same way. In this respect, \citet{Lee2024} and \citet{kolesnikov_paper2} remain the most directly comparable references for the problem considered here, since both explicitly examine the evolution of morphological fractions with redshift within machine-defined classification frameworks.

This comparison matters because parametric and non-parametric methods do not measure the same thing. Parametric methods answer one kind of question: how well does the observed light resemble an idealized profile, or a combination of such profiles? Non-parametric methods answer another: how is the flux distributed in space, independently of whether that distribution resembles a simple analytic form? These are not equivalent questions, especially at high redshift. A galaxy may be well fitted by a low-S\'ersic component and still show a level of concentration, asymmetry, or internal flux organization that places it closer to what a non-parametric system would identify as spheroid-like. Conversely, a compact system may fail to look like a classical elliptical while still being far from a settled disc. The tension between methods should therefore not be treated only as a technical nuisance. It may carry physical information.

In this paper, we revisit this issue using the JWST sample analysed by \citet{Lee2024} over the redshift interval (0.6 < z < 8.0), restricting the analysis to galaxies with stellar masses above  $10^{9}\,{\rm M}_{\odot}$. We apply the methodology developed by \citet{kolesnikov_paper1,kolesnikov_paper2} and compare the resulting non-parametric machine-defined classifications with the parametric classifications obtained by \citet{Lee2024} for the same galaxy population. Rather than treating differences between the two approaches as mere classification noise, we use them to identify where morphology remains stable across methods and where it becomes method-dependent. This comparison allows us to assess whether the disc--spheroid dichotomy remains a meaningful description across cosmic time and to identify the regime in which galaxies may still be too structurally unsettled to admit a unique morphological classification. The goal is therefore not simply to determine which fraction is correct, but to investigate what each method is detecting in systems that are still undergoing structural assembly.

This paper is organized as follows. In Section~2 we describe the sample, data, and selection criteria. In Section~3 we summarize the morphological framework and classification procedure. In Section~4 we explore method stability and internal consistency of the morphological classification.  In Section~5, we present the redshift evolution of the morphological fractions and compare our results with previous work. We also discuss the origin and meaning of the divergence at $z > 2.5$. Section~6 summarizes our conclusions.

\section{Sample selection and data used}
\label{sec:data}

\begin{figure*}
	\includegraphics[width=505px]{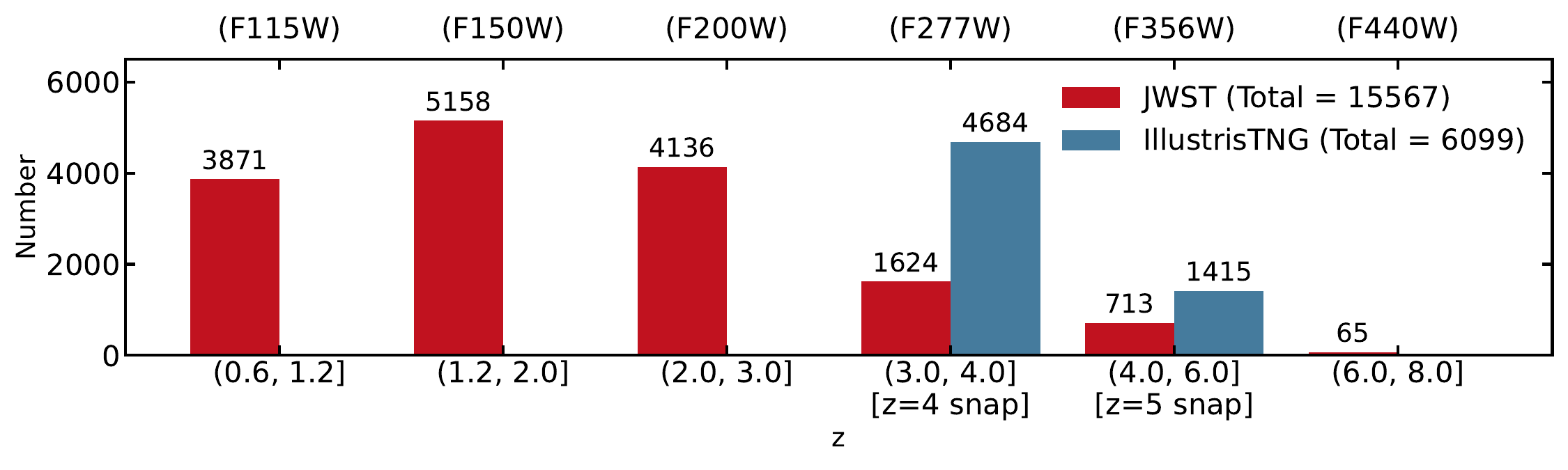}
    \caption{Distribution of galaxies as a function of redshift for the JWST (crimson) and IllustrisTNG (slate blue) samples used in this work. JWST galaxies are grouped into the redshift bins adopted by \citet{Lee2024}, with the corresponding NIRCam filter indicated in parentheses. The numbers above each bar indicate the number of galaxies retained after preprocessing. IllustrisTNG galaxies are available at discrete redshifts ($z=4$ and $5$).}
    \label{fig:data_bins}
\end{figure*}

To investigate the morphological evolution of galaxies across cosmic time, we use the photometric catalogues and physical parameters compiled by \citet{Lee2024}. Our morphological classification differs from theirs, since it is based on the non-parametric hybrid machine-learning framework described in Section~\ref{sec:method_update}. We nevertheless adopt their data reduction, source extraction, and spectral energy distribution (SED) fitting products as the basis for the present analysis.

\subsection{Observational data}

Our sample is drawn from deep JWST/NIRCam imaging in six public extragalactic fields: the North Ecliptic Pole Time-Domain Field \citep[NEP-TDF;][]{Windhorst2022,Windhorst2023}, the Next Generation Deep Extragalactic Exploratory Public survey \citep[NGDEEP;][]{Finkelstein2021,Bagley2024}, the Cosmic Evolution Early Release Science Survey \citep[CEERS;][]{Finkelstein2017,Finkelstein2023}, the Cosmic Evolution Survey \citep[COSMOS;][]{Dunlop2021,Kartaltepe2021,Casey2023}, the UltraDeep Survey \citep[UDS;][]{Dunlop2021}, and SMACS J0723.3-7327 \citep{Pontoppidan2022}. The images were processed and aligned with the \textsc{grizli} pipeline \citep[v6.0;][]{Brammer2023a}, including corrections for instrumental artefacts such as 1/f noise and wisps. The final mosaics have a pixel scale of $0.04''\,\mathrm{pixel}^{-1}$.

\subsection{Photometry and physical parameters}

Source detection and multi-wavelength aperture photometry were performed by \citet{Lee2024} with \textsc{SExtractor} \citep{Bertin1996}, using a dual-mode configuration with cold and hot runs to improve the detection of both bright extended sources and faint compact objects. The combined NIRCam long-wavelength images, F277W + F356W + F444W, were used as the reference detection images.

Photometric redshifts ($z$) and stellar masses ($M_*$) were derived by \citet{Lee2024} through SED fitting with \textsc{eazy-py} \citep{Brammer2008,Brammer2023b}. The catalogue shows good redshift accuracy, with an outlier fraction of $\sim 6\%$ and a normalized median absolute deviation of $\sigma_{\mathrm{NMAD}} = 0.009$ for galaxies with robust fits, defined by $\chi^2_{\mathrm{SED}} < 20$ and $N_{\mathrm{filt}} \ge 6$. We use these quantities for the binning and analysis presented below.

\subsection{Sample selection and rest-frame optical filters}
\label{sec:samplesec}

Following the baseline selection of \citet{Lee2024}, we restrict the parent sample to galaxies in the redshift interval $0.6 < z \le 8.0$ and with stellar masses $M_* > 10^9\,M_\odot$. To ensure that the morphological measurements trace the older stellar distribution, rather than only clumpy young star-forming regions, we select the NIRCam filter that samples the rest-frame optical wavelength range, $\sim 0.5$--$0.7\,\mu\mathrm{m}$, in each redshift bin. The adopted bins and filters are $z \in (0.6,1.2]$ for F115W, $(1.2,2.0]$ for F150W, $(2.0,3.0]$ for F200W, $(3.0,4.0]$ for F277W, $(4.0,6.0]$ for F356W, and $(6.0,8.0]$ for F444W. Each redshift bin is then processed as an independent pipeline run, so that the classification in a given bin is determined only by the galaxies within that bin.

At the cutout and preprocessing stage, our procedure differs from that of \citet{Lee2024}, since it is designed for non-parametric metric extraction. As discussed in \citet{kolesnikov_paper1} and \citet{kolesnikov_paper2}, data preparation is a critical step in the measurement of morphological indices. We create cutouts centered on each galaxy in the sample, with size $KR_p \times KR_p$, where $R_p$ is the Petrosian radius. We use $K=5$ for the images used in the metric extraction and $K=3$ for those used in the Deep Learning (DL) classification (see Section 3.4). Secondary sources within the cutout are masked, producing a segmentation map in which only pixels associated with the primary target are retained (see \citet{kolesnikov_paper2} for details). The larger cutout used for the metric extraction is required for the background subtraction, which needs an extended empty region around the galaxy. For the DL classification, by contrast, only the galaxy image is required.

Because our unsupervised clustering relies on structural indices measured with the MEGG system composed of the second moment of the light distribution ($M_{20}$), Shannon entropy ($E$), the Gini coefficient ($G$), and gradient pattern asymmetry ($G_{2}$), defined in \citet{kolesnikov_paper2}, it is sensitive to image contamination, background residuals, and limited spatial resolution. To make the parameter estimates more robust, we apply an empirical cut based on the number of valid pixels, defined as the non-zero pixels remaining within the segmented mask. After testing different thresholds, we require each galaxy to have at least 100 valid pixels, roughly corresponding to a $10 \times 10$ pixel area. This cut is applied uniformly across all redshift bins. The criterion preserves the morphological diversity of the sample while reducing classification noise from unresolved or poorly segmented systems.

The initial parent sample contains 19,119 galaxies drawn from the data set curated by \citet{Lee2024}. During image preprocessing and cutout generation, we remove 1,722 objects located too close to the detector boundaries or with saturated central pixels, since these features can bias the non-parametric measurements. The minimum number of valid-pixels criterion removes a further 1,830 sources with insufficient spatial information. After these quality and validity cuts, the final sample passed to the morphological metric extraction stage contains 15,567 galaxies (Figure~\ref{fig:data_bins}).

\subsection{IllustrisTNG: sample selection and synthetic images}
\label{sec:data24}

We use the synthetic images generated from simulated galaxies in the TNG50-1 run of the IllustrisTNG simulation suite \citep{Nelson2019,Pillepich2019}.\footnote{Throughout the paper, IllustrisTNG refers only to TNG50-1, the highest-resolution run of the suite.} as a complementary data set to test the stability of our methodology. Instead of assessing internal consistency only by splitting the observational sample, we cross-evaluate models trained on two related but distinct data sets: JWST observations and IllustrisTNG synthetic images. This provides a more stringent test of the hybrid classification framework, because the two data sets are morphologically comparable but not identical in origin, noise properties, or construction.

These images were taken from the public data release of \citet{Costantin2023}. We select simulated galaxies with $M_{\ast} > 10^9\,M_{\odot}$ at snapshot redshifts $4$, and $5$. We also require the total half-stellar mass radius, converted to angular size at the corresponding redshift, to be larger than $1\,{\rm arcsec}$. This selection yields 326 at $z=4$, and 113 at $z=5$. To increase the number of images, each simulated galaxy is rendered from five inclinations ($0^\circ$, $45^\circ$, $90^\circ$, $135^\circ$, and $180^\circ$) and four azimuthal angles ($0^\circ$, $90^\circ$, $180^\circ$, and $270^\circ$). This gives 6,520, and 2,260 images at $z=4$, and $z=5$, respectively. We then apply the same preprocessing steps to the IllustrisTNG data as to the JWST observations. This includes background cleaning, segmentation, and quality assurance flags. An additional $2\,681$ galaxies were removed for not reaching the minimum threshold of 100 "valid pixels". The final sample sizes per redshift snapshot are presented in Fig.~\ref{fig:data_bins}.

The image-generation procedure is described in detail by \citet{Costantin2023}; we summarize it here for completeness. First, noiseless images were generated for each galaxy with the SKIRT radiative transfer code \citep[v9.0;][]{Camps2015,Verstocken2017,Camps2020}, including stellar emission, dust attenuation, and scattering. These noiseless images were then used as input to the \textsc{mirage} tool \citep[v2.2.1;][]{Hilbert2019} to simulate NIRCam observations following the CEERS observing strategy, with a 3-point dither pattern in the CEERS 10 pointing, and including instrumental noise and JWST PSF matching. The simulated images were subsequently processed with the JWST calibration pipeline, v1.4.6, following the reduction strategy adopted for in-flight data. The final images used in our analysis have pixel scale of $0.06\,{\rm arcsec\,pixel^{-1}}$ for F356W.



\section{Morphological measurements and classification}\label{sec:method_update}

In this section, we summarize the hybrid unsupervised--supervised framework used to obtain the morphological classifications. A detailed description of the pipeline is given in \citet{kolesnikov_paper1} and \citet{kolesnikov_paper2}.

The classification procedure has four main components: image cleaning and segmentation, morphological metric extraction with the MEGG system, unsupervised clustering for pseudo-labeling, and convolutional neural network (CNN) training and inference. The first and last components are mainly procedural, whereas the second and third define the morphological classification itself.The MEGG system provides the structural parameter space used for the classification. Each metric is sensitive to a different aspect of the light distribution, and each can contribute to the separation between spheroidal and disc systems. Their combination in a four-dimensional space reduces the ambiguity associated with any single diagnostic.

We apply Self-Organizing Map (SOM) \citep{kohonen1990self,kohonen1982analysis,kohonen1982self} to the four-dimensional MEGG parameter space, identifying regions associated with distinct structural properties. In this application, the SOM tends to place the most structurally distinct systems in opposite regions of the map. These regions isolate the clearest examples of the two morphological classes: smooth, concentrated systems on one side, and extended, structured disc systems on the other. We identify these clusters and assign labels according to their metric values. Clusters with high $E$, $G_2$, and $M_{20}$, together with low $G$, are assigned the \textit{Disc} label. Conversely, clusters with low $E$, $G_2$, and $M_{20}$, together with high $G$, are assigned the \textit{Spheroidal} label. We refer to these well-separated groups as \textit{prominent clusters}. They provide the high-purity training set used in the final supervised CNN stage.

To limit redshift-dependent mixing between galaxies from different cosmic epochs, we treat each redshift bin shown in Figure~\ref{fig:data_bins} as an independent data set. The classification pipeline is run separately in each bin, so that the CNN models are trained and applied only to galaxies within the same redshift interval, unless explicitly stated otherwise.

\subsection{Image preparation and segmentation}

The processing begins with the extraction of an image cutout for each target galaxy. To obtain a consistent structural representation, the stamp size is set to $5R_p \times 5R_p$. The cutout is then cleaned to remove secondary objects and background sources, reducing possible biases in the subsequent parameter estimation. The cleaned image is segmented to produce a morphological mask in which only the pixels associated with the primary galaxy are retained.

During this preprocessing stage, we apply quality-control criteria to identify and remove compromised images. Galaxies are excluded if they are too close to the detector boundaries, if a significantly brighter object lies within $2R_p$, or if the central pixels of the target galaxy are saturated. The result is a sample of validated, cleaned, and segmented galaxy images ready for morphological metric extraction.

\subsection{Morphological metric extraction}

\begin{figure*}
	\includegraphics[width=505px]{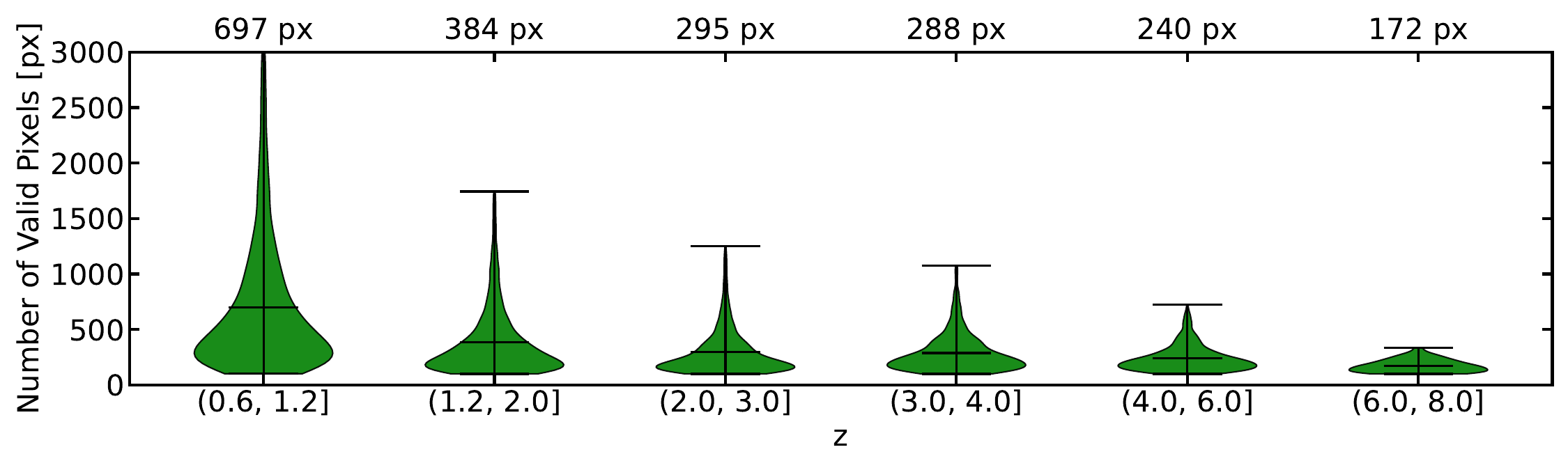}
    \caption{Average number of valid pixels per galaxy as a function of redshift represented as violin plot. Only galaxies with at least 100 valid non-zero pixels within the segmentation mask are included in the analysis (Section~\ref{sec:data}). Numbers above each bin correspond to average of that bin. Distribution was filtered to remove 3$\sigma$ values in order not to overstretch the plot.}
    \label{fig:valid_px_over_z}
\end{figure*}

We extract the MEGG morphological indices from the cleaned and segmented cutouts. The full definitions and implementation details are given in \citet{kolesnikov_paper1} and \citet{kolesnikov_paper2}; here we summarize only the aspects relevant for the present analysis. The MEGG system combines four non-parametric quantities: the second moment of light, $M_{20}$; Shannon entropy, $E$; the Gini coefficient, $G$; and the second gradient moment, $G_2$.

The $M_{20}$ index measures the spatial distribution of the brightest 20 per cent of the galaxy flux \citep{lotz2004new}. It is computed as the normalized second-order moment of the pixels that contain this fraction of the total light, measured with respect to the galaxy center. Low values of $M_{20}$ indicate that the brightest regions are centrally concentrated, as expected for smooth spheroidal systems. Higher values indicate that the brightest regions are more spatially extended, a configuration more typical of disc-like or structured galaxies.

The Shannon entropy, $E$, measures the degree of heterogeneity in the pixel-value distribution \citep{bishop2007pattern}. In this context, low entropy corresponds to smoother and more regular light distributions, while high entropy indicates a larger diversity of pixel values and a more complex internal structure. Entropy therefore provides information that is not reducible to concentration alone.

The Gini coefficient, $G$, quantifies the inequality of the flux distribution among the pixels associated with the galaxy \citep{abraham2003,lotz2004new}. High values indicate that most of the light is concentrated in a small fraction of pixels, independently of where these pixels are located. Low values correspond to a more even distribution of light across the galaxy. In the MEGG space, $G$ is therefore mainly sensitive to flux concentration, while remaining complementary to $M_{20}$ because it does not depend explicitly on the spatial position of the brightest pixels.

The $G_2$ index is based on Gradient Pattern Analysis and measures asymmetries in the gradient field of the galaxy image \citep{rosa2018gradient}. It compares the modulus and phase of vectors located symmetrically with respect to the galaxy center. Symmetric vector pairs are removed, and the remaining asymmetric components are used to quantify perturbations in the flux field. Low $G_2$ values are associated with more symmetric and regular gradient fields, while high values indicate stronger asymmetries and internal structure.

Taken together, these four indices define the MEGG parameter space used in the subsequent classification. The aim is not to assign morphology from any single metric, but to combine concentration, flux inequality, spatial extension, information content, and gradient asymmetry into a single non-parametric description of galaxy structure.

A limitation of non-parametric morphology at high redshift is the amount of spatial information available for each galaxy. Morphological metrics are numerical descriptions of the flux distribution, and their interpretation depends on whether that distribution is sufficiently resolved. In our case, the relevant quantity is the number of valid pixels, defined as the non-zero pixels associated with the galaxy within its segmentation mask. This number decreases toward higher redshift. The effect reflects the combination of finite angular resolution, cosmological surface-brightness dimming, and the smaller apparent sizes of distant galaxies.

To reduce classification noise from unresolved or poorly segmented systems, we require each galaxy to have at least 100 valid pixels, approximately equivalent to a $10 \times 10$ pixel area. This cut removes cases in which the morphological indices would be dominated by insufficient spatial sampling rather than by the intrinsic light distribution of the galaxy. Figure~\ref{fig:valid_px_over_z} shows this effect, with the average number of valid pixels decreasing systematically with redshift across the different morphological classes.

\subsection{SOM and prominent clusters}

After the morphological metrics are measured, we apply SOM to the four-dimensional MEGG parameter space. The SOM projects this space onto a two-dimensional grid while preserving, as much as possible, the neighborhood relations among the input vectors. We use the \texttt{SOMbrero} package, which provides the tools needed to control the fitting and clustering procedure \citep{villa2017stochastic,vialaneix2022package}. At this stage, the algorithm is blind: it uses only the measured morphological indices and has no prior information about the visual or physical class of each galaxy.

The goal at this point is not to classify the full galaxy sample. A complete classification at this stage would be uncertain, because the different classes still overlap within MEGG parameter space. We therefore use the SOM to identify only the clearest regions of this space, where the structural properties are most distinct and the morphological interpretation is least ambiguous.

\begin{figure}
    \centering
    \includegraphics[width=\columnwidth]{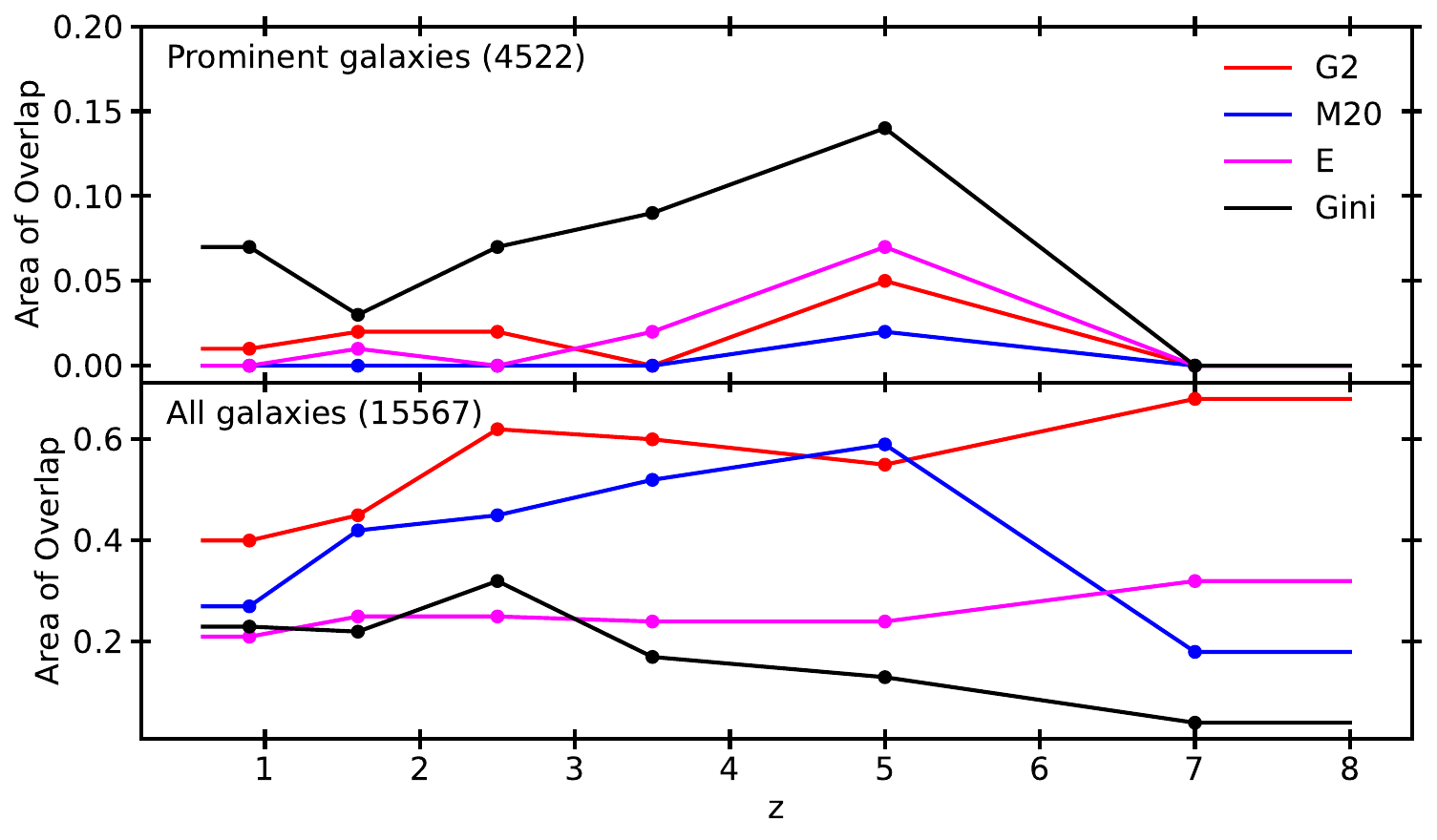}
    \caption{Overlapping area between the distributions of spheroidal and disc galaxies for each morphological metric. Top panels show galaxies selected from the prominent clusters, while bottom panels show the complete galaxy sample. The y-axis displays the fractional overlap area between the distributions \citep{kolesnikov_paper1}. The morphological labels correspond to the final classifications assigned by the pipeline. This figure should be viewed together with Figures~ \ref{fig:metrics}, \ref{fig:priminents_histograms}, and \ref{fig:all_histograms} }
    \label{fig:ao_separation}
\end{figure}

\texttt{SOMbrero} organizes the MEGG parameter space into \textit{clusters}, which are in turn grouped into broader \textit{superclusters}. In our application, the optimized partition separates the MEGG space into two \textit{superclusters}. This division is not imposed by hand; it follows from the within-cluster sum of squares (WCSS) optimization discussed in Sections~3.2 and 3.3 of \citet{kolesnikov_paper2}. The SOM grid defines the set of clusters, while the WCSS-based optimization and the metric thresholds determine which of these clusters occupy the most distinct individual clusters of the parameter space.  These selected individual clusters define the \textit{prominent clusters} used for label assignment. This was shown explicitly with a labelled sample in Figure~6 of \citet{kolesnikov_paper1}. High $G_2$, $M_{20}$, and $E$, together with low $G$, identify the region associated with disc systems. Spheroidal galaxies occupy the opposite region of the MEGG space, with low $G_2$, $M_{20}$, and $E$, and high $G$. This selection is deliberately restrictive. It produces a high-purity subset that is smaller than the parent sample, typically retaining between 20 and 50 per cent of the original data. Figure~\ref{fig:ao_separation} shows the overlap area between the spheroidal and disc distributions for each morphological metric, considering only the prominent clusters. The small overlap indicates that these clusters define a high-purity training set, with the MEGG parameters separating the two morphological classes effectively.

The outcome of this procedure is therefore a pseudo-labelled catalogue built from the most distinct regions of the MEGG space. This subset is used as the training set for the CNN in the supervised part of the hybrid method. In this sense, the SOM does not replace the final classifier; it defines the physically motivated labels on which the hybrid unsupervised--supervised method is based.

\begin{figure}
	\includegraphics[width=\columnwidth]{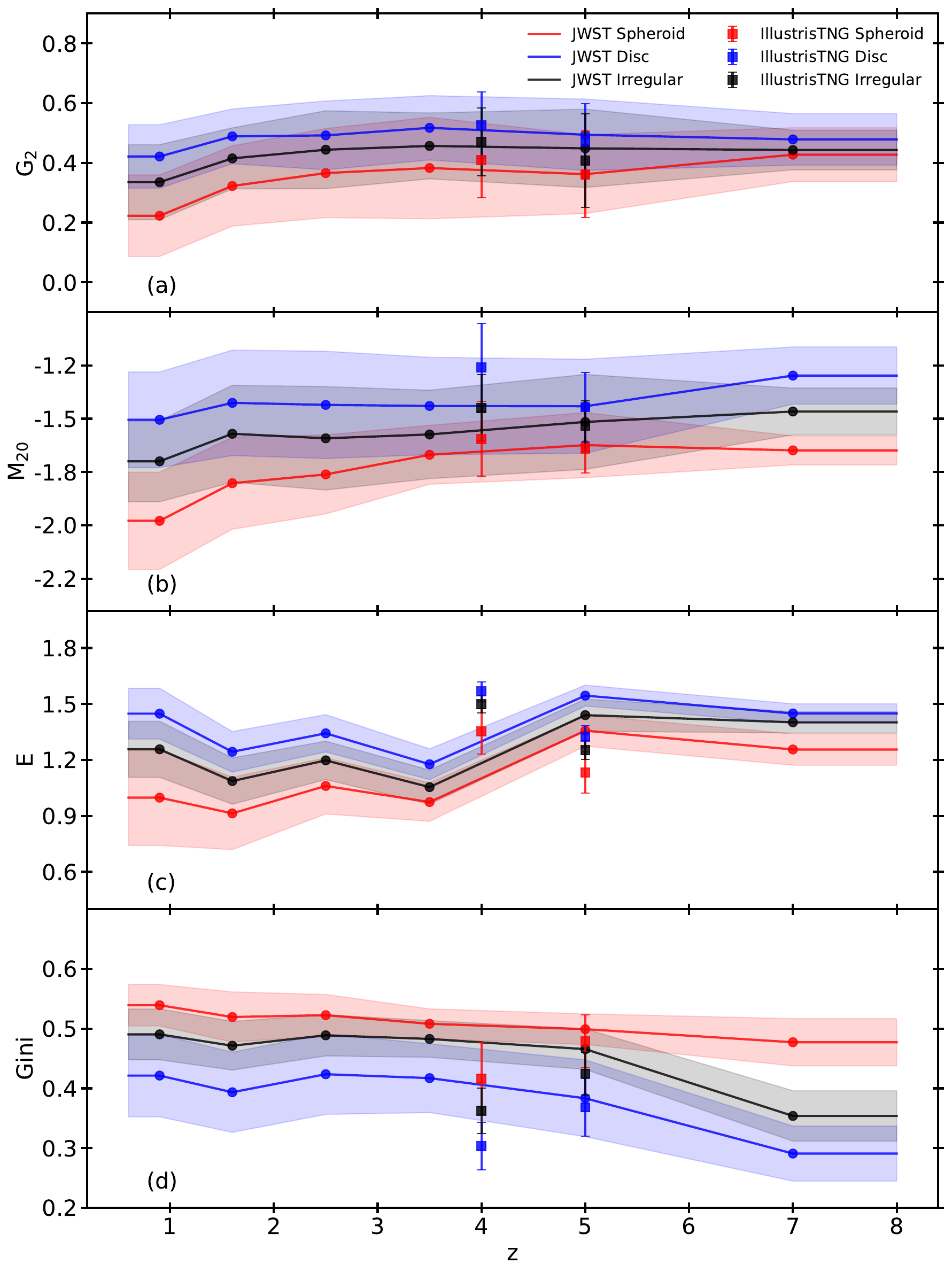}
    \caption{Distribution of the non-parametric morphological metrics for the classified galaxies. Results are shown for the Disc, Spheroidal, and Irregular classes in the JWST and IllustrisTNG samples, separated by redshift bin. The number of galaxies in each bin corresponds to the values reported in Figure \ref{fig:data_bins}. This figure should be viewed in conjunction with Figure \ref{fig:ao_separation}.}
    \label{fig:metrics}
\end{figure}

\subsection{CNN training and supervised classification}

We use \texttt{TensorFlow} as the DL framework for the supervised classification stage \citep{abadi2016tensorflow}. Although the non-parametric indices are measured from cleaned and segmented images, the CNN is trained on the original image cutouts. For each redshift bin, these cutouts are taken from the NIRCam filter selected to sample the rest-frame optical wavelength range, as described in Section~\ref{sec:samplesec}. The FITS cutouts are then converted to PNG format using the \texttt{Trilogy} package \citep{coe2012clash}.

As discussed above, the unsupervised stage is not used to classify the full galaxy sample, because the MEGG parameter space contains substantial overlap between classes. Instead, we use only the most prominent and least ambiguous objects as the pseudo-labelled training set for the CNN. The purpose of the CNN is to learn the image-level structural features associated with each morphological class. Since some uncertainty may remain in the pseudo-labels inherited from the SOM stage, we use an ensemble approach to reduce sensitivity to any particular training subset and to limit overfitting.

For each redshift bin, we train an ensemble of 100 independent CNN models. Each model is trained on a different random split of the training and validation sets, introducing controlled variance into the classification procedure. We use a binary cross-entropy loss function, and the network returns a continuous probability $P$ between 0 and 1 for each galaxy, where larger values correspond to the spheroidal class. The final probability is obtained by averaging the predictions of the 100 models. We then apply confidence thresholds to this averaged probability: galaxies with $P < 0.1$ are classified as discs, while those with $P > 0.9$ are classified as spheroids. Galaxies with intermediate probabilities, $0.2 < P < 0.8$, are assigned to the irregular class. Objects outside these confidence intervals are not assigned a final morphological label and are excluded from the subsequent analysis. 

This ensemble strategy also provides an empirical estimate of classification stability. If the SOM pseudo-labels define physically coherent regions of the MEGG space, the dispersion among the 100 CNN predictions should remain small. In that case, the network is not learning the peculiarities of one training subset, but structural features that are stable across different realizations of the prominent-cluster sample.

After training, the CNN ensemble is applied to the full data set, including the galaxies not labelled during the unsupervised stage. The final morphological label of each galaxy is therefore determined by the averaged CNN prediction. At the end of this stage, approximately 91 per cent of the initial data set is classified as disc, spheroidal, or irregular. This final catalogue is used in the evolutionary analysis presented in the following sections.

\subsection{Morphological separability in the MEGG space}

An important diagnostic of the classification procedure is how well the MEGG parameters separate spheroidal and disc systems at different stages of the method. Figure~\ref{fig:ao_separation} shows the overlap area between the spheroidal and disc distributions for each morphological metric as a function of redshift. The top panel considers only the prominent clusters selected by the SOM, while the bottom one shows the same measurement after the CNN classification is applied to the full sample.

As expected, the prominent clusters show the smallest overlap, since they are selected precisely from the most distinct regions of the MEGG space. This small overlap indicates that these clusters define a high-purity training set, with the MEGG parameters separating the two morphological classes effectively. The full sample shows larger overlap because it includes galaxies with intermediate or less regular structures. This increase does not invalidate the classification. It shows that the CNN extends the labels from the high-purity regions to the broader galaxy population, where the boundaries between spheroids and discs are intrinsically less sharp.

This behaviour can be seen more directly by examining the distributions of the individual MEGG metrics for the three final classes. Figure~\ref{fig:metrics} shows that, even after the CNN is applied to the full sample and the overlap increases relative to the prominent clusters, the expected ordering of the metrics is largely preserved. In the JWST sample, discs occupy the region of higher $G_2$, higher $M_{20}$, and higher $E$, while spheroids show lower values of these three indices. The Gini coefficient follows the opposite trend, with spheroids showing the highest values and discs the lowest. Irregular galaxies generally lie between the two main classes, as expected for systems with mixed or less regular structure. The separation is not constant with redshift, and it becomes less sharp at the highest redshifts, especially for Gini, where the disc and irregular loci move closer together. The IllustrisTNG points broadly follow the same ordering in the overlapping redshift range, although with offsets, particularly in $E$ and Gini. Thus, even when the full sample does not retain the purity of the prominent clusters used for training, the MEGG space still preserves a consistent morphological ranking in both observed and synthetic data.

This diagnostic is also useful for interpreting the comparison with the independent classifications of \citet{Lee2024}, discussed in Section~\ref{sec:fractions_comp}. Their classification is parametric and follows the morphology scheme introduced for the Horizon Run 5 (HR5) simulation by \citet{Park2022}. In the HR5 analysis, galaxies were separated into discs, spheroids, and irregulars using asymmetry and the S\'ersic structure of the stellar-mass distribution. \citet{Lee2024} applied the analogous criteria to JWST images by fitting a single S\'ersic model to the two-dimensional light distribution and by measuring both asymmetry and the residual flux fraction. Regular, well-fitted galaxies with $A<0.32$ and $\mathrm{RFF}<0.5$ were divided according to S\'ersic index, with discs defined by $n<1.5$ and spheroids by $n>1.5$, while galaxies with $A>0.32$ or $\mathrm{RFF}>0.5$ were classified as irregulars. Thus, the comparison below contrasts two different operational definitions of morphology: our non-parametric MEGG-based classification and a parametric, HR5-motivated classification based on S\'ersic structure, asymmetry, and residual light.

\section{Structural Robustness of the Morphological Classification} \label{sec:method_stability}

The classifications presented in this work do not have an external ground truth. Their reliability must therefore be assessed through consistency tests. In this section, we use synthetic images from IllustrisTNG to evaluate whether the inferred morphological labels remain stable under two independent conditions. First, we compare classifications obtained from models trained on observational and synthetic data. Second, we examine whether different projections of the same simulated galaxy receive the same label. These tests do not establish the correct morphology of individual galaxies. Instead, they evaluate whether the classification is driven by intrinsic structural properties rather than by a particular training set or viewing angle.

\subsection{Cross-Domain Consistency Between Observed and Synthetic Galaxies}
\label{sec:jwst_vs_illustris}

\begin{figure}
    \centering
    \includegraphics[width=\columnwidth]{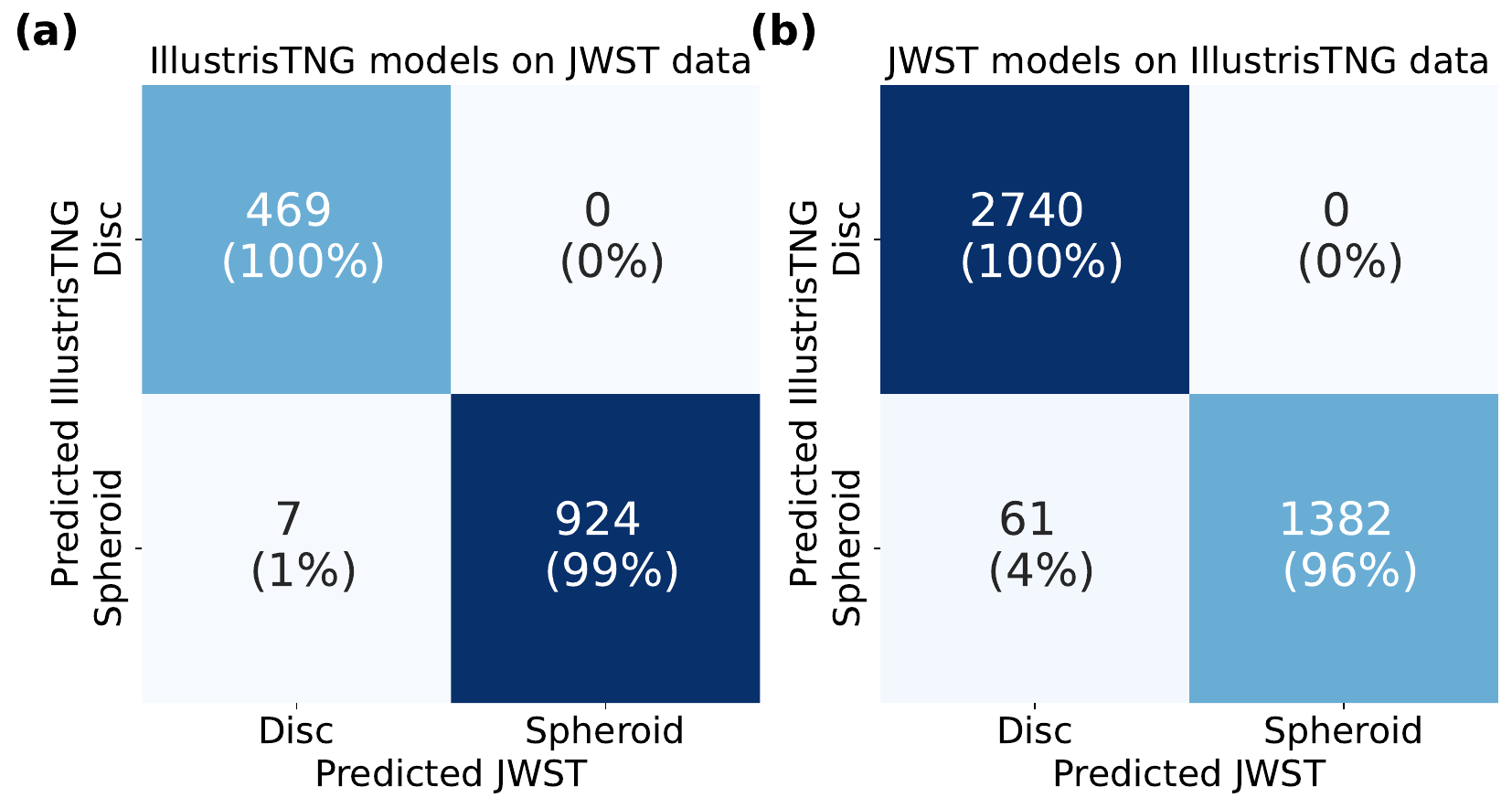}
    \caption{Cross-domain comparison between the JWST and IllustrisTNG classifications.
    Panels (a) and (b) show the confusion matrices obtained from the exchange of model
    ensembles between the two data sets. Panel (a) corresponds to IllustrisTNG-trained
    models applied to the JWST sample, while panel (b) corresponds to JWST-trained
    models applied to the IllustrisTNG sample.}
    \label{fig:illustris_jwst}
\end{figure}

At the redshifts explored in this work, reliable reference morphological classifications are not available. Unlike studies in the nearby Universe, where visual classifications can be used as an external benchmark, the structural complexity and observational limitations of high-redshift galaxies preclude the definition of a robust ground truth. We therefore assess the consistency of the method using the overlap between the JWST and IllustrisTNG samples. As shown in Figure~\ref{fig:data_bins}, two redshift intervals contain both observed and synthetic galaxies. The complexity of this evaluation is notable: rather than simply splitting a single observational data set to train two identical pipeline runs, we are actively cross-evaluating models trained on structurally analogous, yet fundamentally distinct, observational and synthetic data sets. We frame our validation around a core hypothesis: for the learned
features to be deemed physically consistent and valid, models optimized on one data set must reliably classify morphologically analogous galaxies in an independent data set. This allows us to test whether the same morphological classes are recovered when models trained on one data set are applied to the other. The procedure adopted in this comparison is summarized below:

\begin{enumerate}
    \item Observational baseline (\textit{JWST}). We train the final CNN ensemble on the \textit{JWST} data set following the standard pipeline procedure. The classification is performed independently in each redshift bin shown in Figure~\ref{fig:data_bins}. The resulting labels for all 15,567 galaxies are adopted as the reference classification for the observed sample.
    
    \item Synthetic baseline (IllustrisTNG). We train an independent CNN ensemble on the IllustrisTNG data set following the same pipeline procedure. The classification is performed independently in each redshift bin shown in Figure~\ref{fig:data_bins}, using a dedicated model ensemble. The resulting labels for all 6,099 galaxies are adopted as the reference classification for the synthetic sample.
    
    \item Observational-to-synthetic inference. We apply the models trained on the \textit{JWST} data to the IllustrisTNG galaxies. To minimise redshift-dependent effects, comparisons are restricted to corresponding redshift intervals. Specifically, the \textit{JWST} bins $(3 < z \leq 4)$, and $(4 < z \leq 6)$ are matched to the IllustrisTNG snapshots at z=4, and z=5, respectively. The resulting classifications are then compared with the reference labels obtained from the IllustrisTNG models.
    
    \item Synthetic-to-observational inference. We apply the models trained on the IllustrisTNG data to the \textit{JWST} galaxies. As in the previous test, comparisons are restricted to corresponding redshift intervals. The resulting classifications are then compared with the reference labels obtained from the \textit{JWST} models.
    
    \item Consistency criterion. If the independent ensembles learn the same structural information, the classifications should remain stable when the training and inference data sets are exchanged. A high level of agreement would indicate that the morphological separation is not tied to a specific data set, but to structural features shared by the observed and synthetic images.
\end{enumerate}

The results of the cross-domain comparison are shown in Figure~\ref{fig:illustris_jwst}. About \textbf{95} per cent of the galaxies retain the same morphological label regardless of which data set was used to train the model ensemble. This level of agreement indicates that the unsupervised selection of prominent clusters and the subsequent CNN classification recover the same underlying morphological classes in both observational and synthetic data. Furthermore, the confusion matrices of Figure~\ref{fig:illustris_jwst} show that models trained on the observational sample achieve a similar performance as those trained on the synthetic sample. We interpret this symmetry as a consequence of the fundamental similarity of synthetic data to the observational, and as a result, models trained on either dataset generalize effectively when applied to either images.

\subsection{Morphological Robustness Under Multiple Lines of Sight}

In this subsection, we test the sensitivity of the classification to projection effects using the IllustrisTNG synthetic sample. As described in Subsection~\ref{sec:data24}, each galaxy is rendered under 20 independent lines of sight, generated from four azimuthal angles and five inclinations. Each projection is treated as an independent input to the pipeline and receives its own morphological classification.

For each galaxy, we define a majority-vote label (\textit{mvl}) from the classifications of all valid projections of the same galaxy. The stability of the classification is measured as the fraction of projections that recover the \textit{mvl}. A galaxy for which all projections receive the same label has a stability of 100 per cent, whereas a galaxy for which only half of the projections agree with the \textit{mvl} has a stability of 50 per cent. Our analysis is restricted to the three primary classes: spheroid, disc, and irregular. To ensure a reliable majority vote, we include only galaxies with at least five valid projections after the quality-control procedures of the pipeline (see Section \ref{sec:method_update}). We perform this test for both the JWST-trained and IllustrisTNG-trained model ensembles, allowing a direct comparison of their sensitivity to viewing angle.

We detail the performance for both classification ensembles, namely the inference of synthetic snapshots using models trained on observational data, and models trained on synthetic data, following the cross-domain comparison described in Section~\ref{sec:jwst_vs_illustris}. The models demonstrate a mean stability of 82 per cent for the JWST-trained ensembles and 81 per cent for the IllustrisTNG-trained ensembles, corresponding to roughly 17 consistent labels out of 20 possible viewing angles. This high degree of stability holds true when analysed as a function of both galaxy size and redshift. Critically, the JWST-trained models exhibit similar consistency across the entire parameter space. This corroborates the findings of Section~\ref{sec:jwst_vs_illustris} and substantiates the claim that the synthetic data is structurally analogous to natural observations; specifically, it inherently contains the fundamental feature set required to accurately classify observational data, even without being explicitly utilized during the model's training phase.

The stability measured across multiple lines of sight is consistent with the cross-domain comparison presented in the previous subsection. In both cases, most galaxies retain the same morphological classification, indicating that the inferred labels are largely insensitive to viewing angle and to the choice of training data set. The similar performance of the JWST-trained and IllustrisTNG-trained ensembles further supports the structural similarity between the observed and synthetic images.




\section {Results}

Galaxy morphology is not a fixed observational category; it depends on the information carried by the image. In the nearby Universe, visual inspection is often sufficient to separate discs and spheroids and to define training sets with limited ambiguity. At high redshift this condition no longer holds. Surface-brightness dimming, poorer physical resolution, lower signal-to-noise ratio, and the increasing frequency of clumpy or asymmetric structures make visual labels less stable and more dependent on the classifier. For this reason, the analysis presented here does not rely on training sets defined by visual morphology. The aim is to obtain a classification whose reference frame is set by measured structural properties rather than by subjective labels.

In \citet{kolesnikov_paper2}, we show that the non-parametric, unsupervised scheme exposes limitations of visual classification and remains consistent with the parametric results of \citet{Lee2024} up to $z = 2.5$. Here we apply the same comparison over the full redshift range covered by \citet{Lee2024}, reaching $z = 8$. The comparison is therefore no longer restricted to the regime where classical morphologies are expected to be reasonably well established. It also probes the epoch where irregularity, compactness, and low surface-brightness structure make the separation between discs and spheroids intrinsically more uncertain.

\subsection {Observed and simulated morphological fractions across redshift}\label{sec:fractions_comp}

\begin{figure}
    \centering
    \includegraphics[width=\columnwidth]{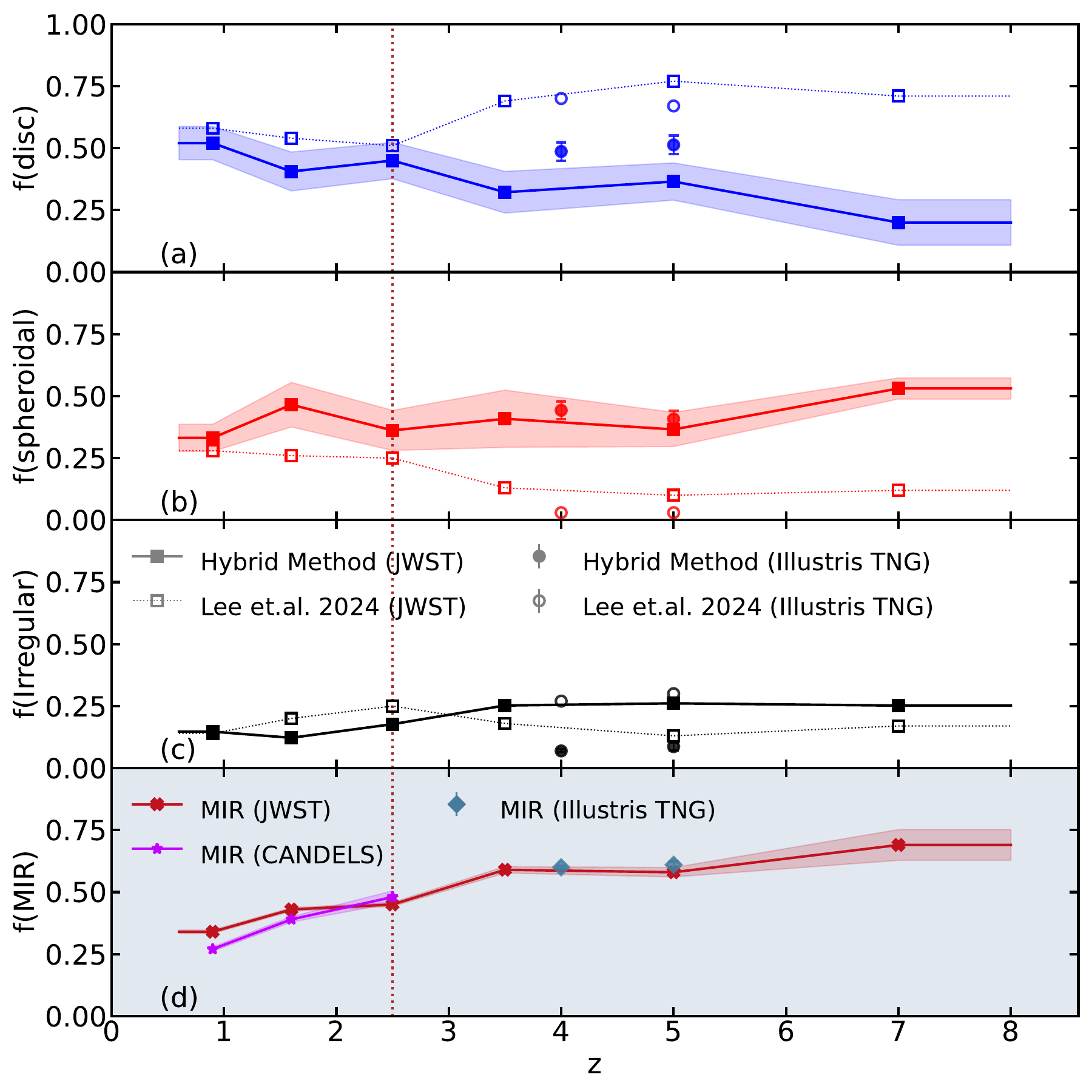}
    \caption{Comparison of the morphological fractions obtained with the classification schemes of \citet{Lee2024} and \citet{kolesnikov_paper1}. Panels (a)-(c) show the fractions of spheroidal, disc, and irregular galaxies as a function of redshift for the JWST and IllustrisTNG samples. Panel (d) shows the redshift evolution of the Morphological Indeterminacy Regime fraction, $F_{\rm MIR}$, defined as the fraction of galaxies outside the region of robust agreement between the two classification schemes. The vertical dotted line marks the redshift above which we compare our results with those of \citet{Lee2024}.}
    \label{fig:fractions}
\end{figure}

Figure~\ref{fig:fractions} shows the redshift evolution of the disc, spheroidal, and irregular fractions. We first compare our JWST hybrid classification with the parametric classification of \citet{Lee2024}. Below $z \simeq 2.5$, the two estimates provide a broadly consistent view of the population. The agreement is not exact, but the differences remain moderate. In this redshift interval, both methods recover a substantial disc population, a significant spheroidal component, and a smaller irregular fraction. Defining $\Delta f = f_{\rm hybrid} - f_{\rm Lee}$, the hybrid classification gives disc fractions lower by $\langle \Delta f_{\rm disc} \rangle \simeq -0.08$, spheroidal fractions higher by $\langle \Delta f_{\rm spheroid} \rangle \simeq 0.13$, and irregular fractions comparable within $\langle \Delta f_{\rm irregular} \rangle \simeq -0.05$. Thus, in the regime where classical morphologies are expected to be more stable, the non-parametric and parametric classifications do not give identical fractions, but they preserve the same global ordering of the morphological classes.

The behaviour changes above $z \simeq 2.5$. In this regime, the classification of \citet{Lee2024} remains strongly disc dominated, whereas our hybrid classification gives a lower disc fraction and a much larger spheroidal fraction. For the $z > 2.5$ domain, the approximate offsets are $\langle \Delta f_{\rm disc} \rangle \simeq -0.43$, $\langle \Delta f_{\rm spheroid} \rangle \simeq 0.32$, and $\langle \Delta f_{\rm irregular} \rangle \simeq 0.09$. The main discrepancy is therefore not driven by the irregular class. The irregular fraction is higher in our classification, but only at the level of $\sim10$ per cent. The dominant effect is a redistribution between discs and spheroids. A large fraction of the systems classified as discs by the parametric approach is assigned to the spheroidal class by the non-parametric method. This difference appears exactly where morphology is expected to become least stable. At $z > 2.5$, galaxies are generally more compact, more irregular, and more affected by low surface-brightness structure and limited spatial resolution. In this regime, a parametric classification based on the dominant light profile can favour disc-like solutions even when the non-parametric structure does not support a settled disc interpretation. The divergence between the two methods may be indicating that the mapping between observed light distribution and classical morphological labels becomes increasingly method dependent at high redshift.

To assess the robustness of these trends, we also quantify the field-to-field variation of the disc and spheroid fractions. For each redshift bin, we compute the weighted mean fraction and the corresponding weighted scatter among the three fields (CEERS, COSMOS and UDS), using the number of galaxies in the bin as weights. The resulting relative scatter, measured as the ratio between the weighted scatter and the weighted mean, increases with redshift for both morphological classes. On average, the field-to-field variation rises from $\sim 13$ per cent at $z\sim 1$ to $\sim 32$ per cent at $z\sim 5$. This increase is seen for both discs and spheroids, indicating that the uncertainty is not restricted to a single morphological class. The growing scatter at high redshift is therefore consistent with the picture that morphological classifications become progressively less determinate in the early Universe. This point is developed further in the next subsection, where we discuss the Morphological Indeterminacy Regime.

The comparison with IllustrisTNG provides an independent test of the robustness of both methods been used here. At the three available snapshot redshifts, we apply both approaches to the simulated images in the same way as for the JWST data. We find the mean difference between the non-parametric and parametric methods to be $\langle \Delta f_{\rm disc} \rangle \simeq -0.23$, $\langle \Delta f_{\rm spheroid} \rangle \simeq 0.37$, and $\langle \Delta f_{\rm irregular} \rangle \simeq -0.13$. These offsets reflect the behaviour shown in panels (a), (b), and (c) of Fig.~\ref{fig:fractions}. When applied to the simulated images, the hybrid method reproduces the JWST disc fraction reasonably well and the spheroidal fraction very well, while the irregular fraction differs by roughly $\sim 0.2$. The parametric methodology, in turn, reproduces the JWST disc fraction reasonably well, underestimates the spheroidal fraction by $\sim 0.12$, and gives an irregular fraction that differs by only $\sim 0.15$.

Taken together, the results indicate two separate effects. First, the comparison with \citet{Lee2024} shows that the main methodological divergence above $z \simeq 2.5$ is the exchange between discs and spheroids. Second, the comparison with IllustrisTNG shows that each methodology preserves, in the simulated images, the same behaviour seen in the observed data. In this sense, both methods are internally consistent, but they do not assign the same morphological meaning to the same galaxy population. This sets an indeterminacy in the definition of a unique morphological classification for high-redshift galaxies. The fraction associated with this morphological indeterminacy regime is discussed in the next subsection.

\subsection {Morphological Indeterminacy Regime - MIR}

The divergence between parametric and non-parametric classifications above $z \sim 2.5$ suggests that galaxies enter a regime in which classical morphological categories cease to provide a unique description of structure. The comparison between parametric and non-parametric classifications reveals a systematic difference in the inferred morphological composition of galaxies at high redshift. While the detailed behavior of the individual fractions has been discussed in the previous section, a key result emerges when the two methods are considered together.

Over the interval $2.5 < z < 8$, the non-parametric classification yields only a mild separation between spheroids and discs, with $\rm f_{s} - f_{d} \sim 0.14$. In contrast, the parametric classification implies a much stronger imbalance between the same two classes, with $\rm f_{s} - f_{d} \sim 0.5$. At the same time, the fraction of galaxies classified as irregular remains significant across the same redshift range. In the non-parametric scheme we find $\rm f_{Irr} \sim 0.25$, while the parametric classification yields a somewhat smaller, but still non-negligible, value of $\rm f_{Irr} \sim 0.12$. Together, these results indicate that the high-redshift galaxy population is structurally more heterogeneous than a simple disc--spheroid dichotomy would suggest.

We interpret this behavior as evidence that galaxies in the early Universe occupy a regime in which classical morphological categories are not uniquely defined. Parametric approaches necessarily impose smooth analytic profiles on the observed light distribution and therefore tend to regularize structurally complex systems, assigning them to one of the traditional morphological families. Non-parametric diagnostics, by contrast, respond directly to the spatial distribution of flux and are therefore more sensitive to concentration, clumpiness, and asymmetry. When galaxies are dynamically immature --- compact, clumpy, and still assembling --- the two approaches may therefore probe different aspects of the same system.

\begin{figure*}
    \centering
    \includegraphics[width=505px]{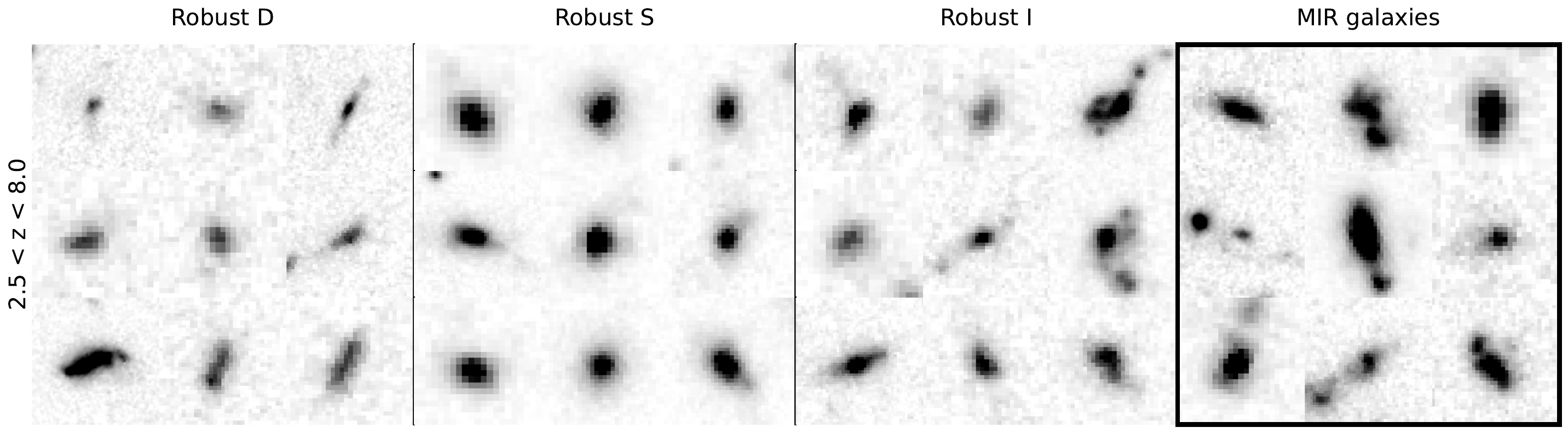}
    \caption{Representative examples of robust disc (D), spheroidal (S), irregular (I), and MIR galaxies in the redshift range (3 < z < 8). The robust classes correspond to galaxies assigned the same morphology by both classification schemes, while MIR galaxies correspond to systems with discrepant classifications.}
    \label{fig:mir_mosaic}
\end{figure*}

We refer to this epoch as the \textit{Morphological Indeterminacy Regime}. In the MIR, different structural diagnostics do not necessarily converge to a unique morphological classification because galaxies have not yet developed the dynamically settled structures characteristic of the Hubble sequence. The disagreement between the methods is therefore not simply methodological; rather, it reflects the fact that galaxy morphology itself is not yet fully defined during this early phase of galaxy assembly. Figure~\ref{fig:mir_mosaic} presents representative examples of each morphological class. Discs and spheroids exhibit the expected contrast in structural appearance, whereas MIR galaxies appear more similar to irregular systems, often displaying a higher degree of fragmentation and substructure.

In this picture, the primordial galaxy population cannot be described adequately by a simple opposition between discs and spheroids alone. A significant fraction of systems are identified as irregular, while another fraction appears to depend explicitly on the diagnostic used. The approximately constant behavior of the morphological fractions observed over $2.5 < z < 8$ suggests that this regime persists over a substantial portion of early cosmic history, during which galaxy structures are still evolving toward the more stable configurations observed at lower redshift. Also, we have to remember that this redshift interval corresponds to only $\sim$1.98 Gyr (14 percent of the age of the Universe). The MIR therefore provides a natural framework for interpreting discrepancies between morphological methods at high redshift. Rather than indicating a failure of the classification techniques, such discrepancies may instead signal that galaxies themselves have not yet reached a stage where their morphology can be uniquely described within the traditional disc--spheroid paradigm.

The comparison between the two classification schemes can be summarized in a set of simple but revealing fractions, for $z > 2.5$. After excluding objects classified as uncertain in the hybrid system, we find that the fraction of galaxies identified as discs by both methods is \(f_{DD}=0.31\pm0.01\), whereas the fraction identified as spheroids by both methods is \(f_{SS}=0.11\pm0.01\). In addition, a smaller fraction, \(f_{II}=0.04\pm0.004\), corresponds to systems classified as irregular by both schemes, representing a joint channel for strongly disturbed or non-standard morphologies. When the two robust agreement terms are added, one obtains \(f_{\rm robust}=f_{DD}+f_{SS}=0.42\pm0.01\), showing that only about \(42\%\) of the sample receives a concordant classification. Including the irregular agreement term raises the common-interpretation fraction only modestly, and the remaining population is quantified by \(F_{\rm MIR}=1-(f_{\rm robust}+f_{\rm Irr})=0.54\pm0.01\), where \(f_{\rm II}=0.04\pm0.004\). This means that roughly \(54\%\) of the galaxies lie outside the region of straightforward morphological agreement. Such a large value is even more difficult to interpret as a minor statistical fluctuation or a secondary systematic effect. Instead, it indicates that more than half of the sample occupies a regime in which morphology is intrinsically method-dependent, with the two schemes responding to different structural aspects of the galaxies. The dominance of \(f_{DD}\) over \(f_{SS}\) remains noteworthy, since the robust disc--disc agreement is still substantially larger than the spheroid--spheroid agreement, suggesting either that disc-like structure is more stably recovered across methods or that spheroidal identification is more sensitive to the particular assumptions built into each classifier. More generally, these numbers strengthen the conclusion that the disagreement between the parametric and non-parametric views of galaxy structure is not confined to a small tail of problematic objects, but instead affects the majority of the sample in this redshift interval. In this sense, \(F_{\rm MIR}\) can be understood as measuring the extent of a mixed-interpretation regime, that is, the domain in which galaxies cannot be assigned a unique morphology independently of the adopted method. This has an important physical implication: once galaxies enter this regime, the classical categories of disc and spheroid cease to function as fully method-invariant descriptors, and any statement about the evolution of morphological fractions must be read with the awareness that a large fraction of systems may not belong cleanly to either class in a universal sense.

The relevance of MIR becomes even clearer when \(F_{\rm MIR}\) is followed as a function of redshift. Panel (d) of Figure~\ref{fig:fractions} shows that the mixed-interpretation fraction is not constant, but increases systematically toward earlier cosmic times. At \(z\lesssim 2.5\), \(F_{\rm MIR}\) remains relatively modest, at the level of \(\sim 35\%\). In this regime, most galaxies can still be assigned a reasonably stable morphological identity across independent classification schemes. The agreement between the HST- and JWST-based estimates of \(F_{\rm MIR}\) in this redshift interval is also important. It shows that the measured fraction is not driven by a particular data set, instrumental configuration, or reduction procedure, but reflects a robust level of method-dependent ambiguity in the classification itself. Toward higher redshift, however, the fraction of systems lacking a method-invariant classification grows steadily, reaching values close to 0.7 in the highest-redshift bin. The trend is therefore not a marginal modulation around a fixed level of disagreement. It points instead to a genuine evolutionary effect: the further back one looks, the larger the fraction of galaxies whose structure cannot be mapped cleanly into the classical disc--spheroid framework. One possible concern is that the increase in \(F_{\rm MIR}\) simply reflects the reduced spatial information available at high redshift. However, Figure~\ref{fig:valid_px_over_z} shows that the average number of valid pixels per galaxy remains relatively stable over the redshift range where MIR is measured, varying by at most \(5\)--\(18\) per cent. The rise of \(F_{\rm MIR}\) is therefore substantially larger than any corresponding change in image sampling, suggesting that the observed trend cannot be explained solely by decreasing galaxy size or resolution effects.

This result strengthens the physical interpretation of MIR. Rather than being a by-product of imperfect classification, MIR appears to trace an epoch in which a large part of the galaxy population is still dynamically young, structurally composite, or only partially relaxed. In this view, the rise of \(F_{\rm MIR}\) with redshift is itself evidence that morphological indeterminacy is a characteristic feature of early galaxy assembly.

\subsection {Stellar mass distributions at $2.5 \le z \le 8.0$}

\begin{figure}
    \centering
    \includegraphics[width=\columnwidth]{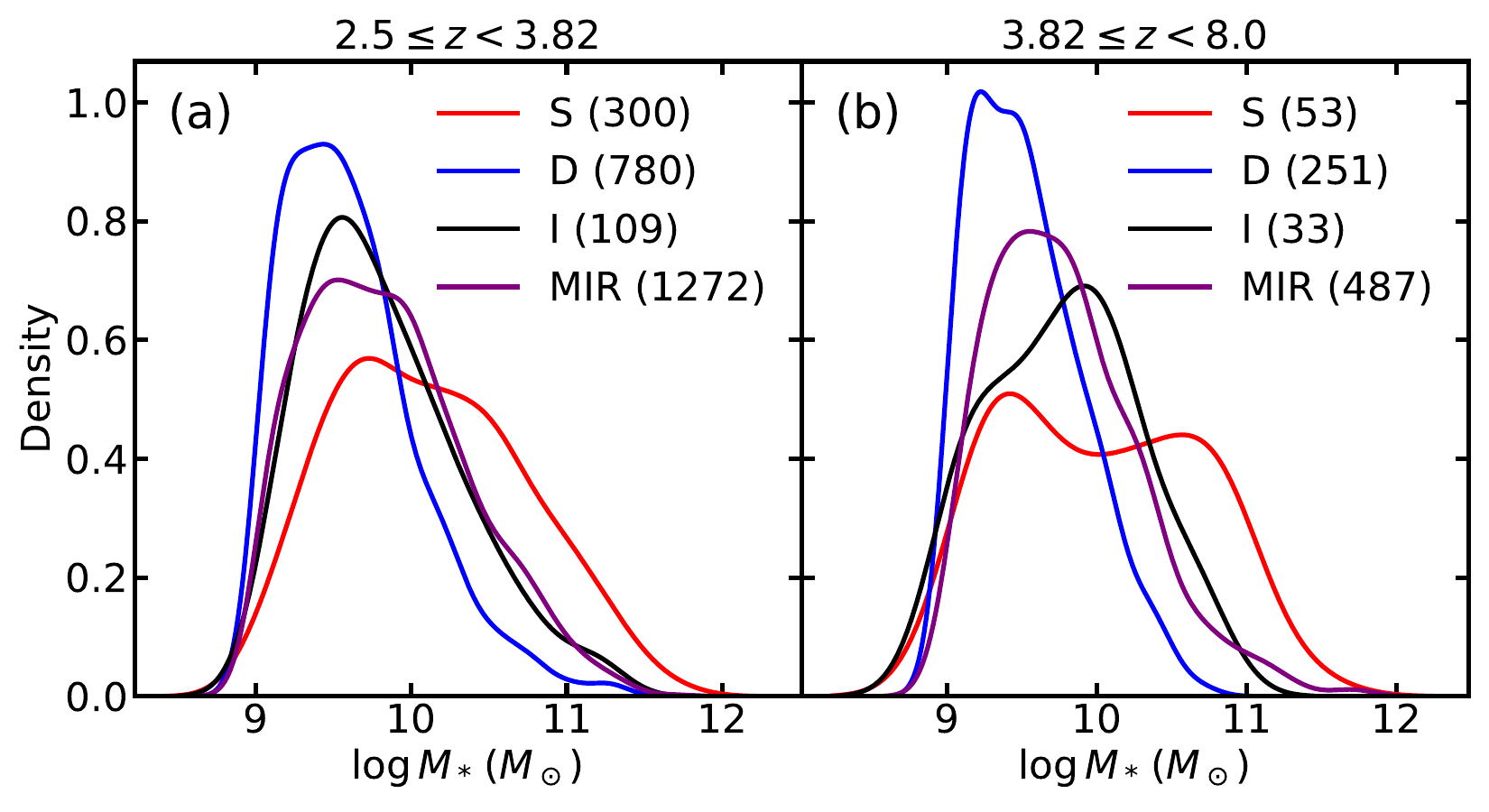}
    \caption{Stellar mass distributions of galaxies in the redshift interval ($2.5 < z < 8.0$), dividing into two redshift intervals that span equal intervals in cosmic time. The two bins each cover approximately 1 Gyr: $2.5 \le z < 3.82$ (panel a) and $3.82 \le z \le 8.0$ (panel b). Spheroidal, disc, and irregular galaxies correspond to the robust classifications obtained from the agreement between the two morphological schemes, while MIR galaxies correspond to systems outside the region of robust agreement. The numbers in parentheses indicate the number of galaxies in each category. Distributions are smoothed using an Epanechnikov kernel.}
    \label{fig:mir_mass}
\end{figure}

Figure~\ref{fig:mir_mass} shows the stellar mass distributions of discs, spheroids, irregulars, and MIR galaxies at $2.5 < z < 8.0$. The sample is divided into two redshift intervals that span equal intervals of cosmic time. Since the full redshift range corresponds to $\sim 2$ Gyr, the two bins each cover approximately 1 Gyr: $2.5 \le z < 3.82$ (panel a) and $3.82 \le z \le 8.0$ (panel b). Despite the short time interval, the mass ordering is already clear. Discs have the lowest median stellar masses ($\sim 10^{9.5},M_{\odot}$), while spheroids have the highest ($\sim 10^{10.1},M_{\odot}$). Irregular and MIR galaxies occupy an intermediate regime, with nearly overlapping distributions and median stellar masses of $\sim 10^{9.7},M_{\odot}$. This ordering suggests that the MIR population is not randomly distributed in stellar mass. Instead, it lies in the mass range where structural assembly is most active, between lower-mass settled discs and more massive spheroidal systems. A further notable feature is the apparent bimodality of the spheroid mass distribution in both panels. This result is consistent with the broader picture emerging from \citet{Sampaio2025}, in which spheroid-dominated galaxies do not form a single homogeneous population.

This result is consistent with a hierarchical interpretation. MIR and irregular galaxies may trace systems in which morphology is still dynamically unsettled, either because of ongoing assembly, interactions, clump migration, or incomplete relaxation. In this sense, MIR is not simply a classification uncertainty, but a population-level signature of structural indeterminacy during the build-up of galaxies. The spheroidal distribution also appears to be bimodal, suggesting that spheroids at these redshifts may not form a single homogeneous population. A lower-mass component may correspond to compact or early-formed spheroidal systems, while the high-mass component may trace the more advanced products of rapid assembly and dynamical relaxation. This will be addressed elsewhere.

\section{Discussion}

Within the $\Lambda$CDM scenario, the main implication of these results is direct. The early Universe does not need to produce galaxies that fall cleanly, and in a method-independent way, into the classical morphological classes defined from the nearby Universe. Hierarchical growth is an assembly process. Galaxies accrete gas, exchange angular momentum, experience interactions and mergers, form stars in compact regions, develop instabilities, and only gradually approach more stable dynamical states. Under these conditions, settled discs and relaxed spheroids should not exhaust the structural diversity of the high-redshift population. The range of $0.50 < F_{\rm MIR} < 0.70$, for $z > 2.5$, is therefore physically significant. It indicates that a large fraction of galaxies occupy a regime in which structure is still being built, rearranged, or only partially relaxed, so that different classification schemes recover different aspects of the same object. This interpretation is consistent with the current JWST literature. \citet{Lee2024} find a disc-dominated population up to high redshift using a parametric classification, whereas \citet{Kartaltepe2023} find a broad diversity of visual morphologies at $z=3$--9, with discs, spheroids, and irregular features all present in significant fractions. \citet{Tohill2024} use an unsupervised approach to identify eleven morphological classes at $z>2$ and argue that conventional classification systems are too restrictive for this regime. \citet{VegaFerrero2024} further show that a substantial fraction of galaxies classified as discs at $z\geq 3$ occupy a region of machine-learning representation space associated, in TNG50, with systems of low stellar specific angular momentum and non-oblate structure, suggesting that the high-redshift disc fraction may be overestimated by traditional classifications. Although the analysis of \citet{Huertas-Company2024} is based on visual morphological classifications, and is therefore subject to some degree of intrinsic subjectivity, it provides a useful external benchmark. Their comparison with \citet{Lee2024} shows that the main differences occur between the disc and irregular classes at $4<z<6$. A similar trend is found when their results are compared with the hybrid classifications discussed here. Moreover, when \citet{Huertas-Company2024} group their sample only into regular and disturbed systems, they find that at $z>3$ the population is dominated by disturbed galaxies, which account for $\sim$90 per cent of the sample. The discrepancy is therefore not a simple disagreement in numbers. It is a tension between taxonomies, methods, and the physical meaning assigned to disc-like, spheroidal, compact, and irregular light distributions. In this sense, $F_{\rm MIR}$ measures the size of an expanded morphological domain in which the observed light distribution has not yet converged to a unique, method-invariant form. From the point of view of $\Lambda$CDM, this is expected. Galaxies are assembled through accretion, instability, interaction, and subsequent settling. The redshift evolution of discs and spheroids may therefore trace not only the relative growth of two canonical families, but also the changing size of this mixed-interpretation population. Morphological disagreement becomes, in this context, an observational signature of hierarchical assembly.

At $z>2.5$, differences between morphological classifications are likely to reflect more than the technical details of the methods. Both the MEGG-based and parametric classifications are applied to rest-frame optical images, and therefore primarily describe the distribution of stellar light. They do not directly trace the gaseous component, which may be more extended, more irregular, and more dynamically unsettled than the stellar body. This distinction is important for the MIR interpretation, because high-redshift galaxies may already contain compact or disc-like stellar structures while still being embedded in larger, actively assembling gaseous systems. Integral-field spectroscopy with the Multi Unit Spectroscopic Explorer (MUSE) on the VLT provides a useful physical context for this picture. In the MUSE HUDF, \citet{Leclercq2017} detect extended Ly$\alpha$ haloes around 145 galaxies at $3 \le z \le 6$ and find that, in about 80 per cent of the reliable cases, the Ly$\alpha$ emission is more extended than the UV continuum, often by factors of $\sim 4$ to $>20$. \citet{Wisotzki2018} further show that this low-surface-brightness Ly$\alpha$ emission around galaxies at $3<z<6$ reaches a projected sky coverage close to 100 per cent. Thus, circumgalactic neutral hydrogen is not a secondary feature of a few systems; it is a widespread component of the high-redshift population. In the context of MIR, this matters because the stellar body used for morphological classification is often only the bright core of a larger gaseous and dynamically active system. A galaxy identified in imaging as a disc or spheroid may therefore represent only one structural component of a system that is still assembling. MIR is then not merely a disagreement between classifiers, but traces a regime in which the stellar light distribution has not yet become a unique descriptor of the physical state of the galaxy.

The more direct kinematic support comes from the diversity of Ly$\alpha$ spectral behavior revealed by deep MUSE spectroscopy. \citet{Leclercq2020} show that Ly$\alpha$ haloes exhibit significant intra-halo variations in line profile, and that the halo component is often broader, spectrally shifted, and less asymmetric than the central component, which already argues against a simple, single-phase, dynamically settled picture. \citet{Vitte2025} extend this point statistically using the MUSE eXtremely Deep Field: from 477 Ly$\alpha$-emitting galaxies at $2.8<z<6.6$, with an unbiased subset of 206, they find that roughly 32--51 per cent show double-peaked Ly$\alpha$ profiles, with peak separations from about 150~km~s$^{-1}$ to nearly 1600~km~s$^{-1}$, while about 20 per cent of the parent sample lies in a complex environment with nearby clumps or companions within 30~kpc. Since Ly$\alpha$ is a resonant line, these signatures cannot be reduced to a one-to-one measurement of intrinsic gas velocity dispersion; they encode scattering, neutral-gas geometry, inflows, outflows, and environmental structure together. But that is precisely why they are so relevant for MIR. They show that a substantial fraction of high-redshift galaxies inhabit a regime in which the emergent line morphology is itself multivalued, environmentally coupled, and sensitive to non-trivial gas dynamics. This is fully consistent with the idea that, above $z \sim 2.5$, many galaxies have not yet converged to the ordered internal states normally associated with mature discs or relaxed spheroids. The Ly$\alpha$ results therefore do not merely accompany MIR; they provide an independent phenomenological basis for it, by showing that the gaseous component of the same high-redshift population is spatially extended, spectrally diverse, and dynamically non-uniform.

In this framework, MIR marks an assembly regime that precedes the epoch of maximum cosmic star formation efficiency. At $z \gtrsim 3$, gas is abundant, but a large fraction of galaxies is not yet organized into long-lived, dynamically coherent structures capable of sustaining extended star formation. The temporal sequence is therefore meaningful: MIR dominates at $z>3$, while the cosmic star formation rate density peaks later, at $z \sim 2$--$2.5$ \citep{MadauDickinson2014}. This suggests that the decline of MIR traces the progressive settling of galaxies into stable configurations, including rotationally supported discs. In this view, the peak of the cosmic star formation history occurs when a critical fraction of galaxies has already moved out of the MIR regime. The subsequent decline at $z \lesssim 2$ then reflects gas depletion and the increasing role of quenching, not a return to morphological indeterminacy.


\section{Summary}\label{section_conclusions}

In this work, we investigate the morphological evolution of galaxies in the $0.6 < z \leq 8.0$ range, utilizing a mass-complete sample ($M_{*} > 10^9 M_{\odot}$) comprising 15,567 JWST observations. To circumvent the limitations and subjective biases inherent in visual inspection, we employ a hybrid unsupervised-supervised machine learning approach grounded in the non-parametric MEGG system ($M_{20}$, Entropy, Gini, and Gradient Pattern Analysis) to categorize this data set into disc, spheroid, and irregular classes. We contrast the morphological labels generated by our data-driven, non-parametric pipeline, with those provided by \cite{Lee2024}, who classify the same data set using a distinct, parametric unsupervised approach. We also measure how these two methods perform on synthetic snapshots images extracted from the IllustrisTNG simulations at redshifts 4, and 5. We summarize the main findings of this work as follows:

\begin{itemize}
    \item While parametric and non-parametric morphological classifications exhibit reasonable agreement at $z < 2.5$, a systematic divergence emerges at higher redshifts. For $z > 2.5$, parametric methods \citep{Lee2024} tend to infer a predominance of disc galaxies, whereas our non-parametric approach yields a significantly higher fraction of spheroidal systems. 
    \item We interpret this high-redshift discrepancy not as a methodological failure, but as the observational imprint of a morphological indeterminacy regime, within which traditional disc and spheroid categories cease to function as universal, method-invariant descriptors, primarily because these systems have not yet converged to the ordered internal states normally associated with mature discs or relaxed spheroids.
        \item The internal consistency of our classification framework is confirmed through rigorous cross-match validation between JWST observations and IllustrisTNG synthetic data. Our models demonstrate an average consistency of $\sim$ 95 per cent when reciprocally applied to the observational and simulated data sets, retaining the same morphological label independent of the training source. This robust agreement indicates that the extracted structural features are physically meaningful and not driven by observational artefacts.
    \item We find that the fraction of this indeterminate class of galaxies, $F_{\rm MIR}$, systematically increases from $\sim 0.5$ at $ z \sim 2.5$ to $\sim$ 0.7 at $z = 8.0$. Consequently, approximately 60 per cent of the high-redshift galaxy population cannot be cleanly assigned to a classical Hubble sequence category, underscoring that morphological indeterminacy is an intrinsic characteristic of galaxy assembly at high $z$.
    \item The prevalence of the MIR supports a hierarchical structure formation scenario in which the early Universe is predominantly populated by dynamically immature, composite systems. The observed morphological indeterminacy serves as an observable signature of hierarchical assembly, reflecting a transitional phase that precedes the emergence of settled, rotationally supported discs and relaxed spheroids. Besides, MIR galaxies preferentially occupy an intermediate stellar mass regime between low-mass discs and massive spheroids, suggesting that the MIR phase is linked to the epoch of most active structural assembly rather than being randomly distributed across stellar mass.
\end{itemize}

In essence, our findings underscore that morphological ambiguity at high redshift constitutes a physical reality of galaxy assembly. Imposing classical low-redshift taxonomies on the early Universe obscures the structural richness of these evolving systems, highlighting the necessity of data-driven, non-parametric approaches in the JWST era.

\section{Acknowledgements}
RRdC and RFF acknowledge the support from FAPESP through the grants 2020/15245-2 and 2025/24207-0, respectively. VMS thanks the Agencia Nacional de Investigacion y Desarrollo (ANID), through the Millennium Science Initiative Program NCN2024-112, and the additional support from ESO grant ORP026/2021. J.H.L. acknowledges the support from the Basic Science Research Program through the National Research Foundation of Korea (NRF) funded by the Ministry of Education (No. RS-2024-00452816). LASL is supported by CNPq grants 402281/2026-8 and 326437/2026-6.

This research made use of several open-source Python packages, and we are deeply grateful to the developers and maintainers of these tools. Specifically, this work relied on \texttt{NumPy} \cite{harris2020array} for efficient array programming and numerical computation, \texttt{SciPy} \cite{virtanen2020scipy} for foundational scientific algorithms, and \texttt{pandas} \cite{mckinney2010data} for data structures and data manipulation. 

Data visualization and figures were generated using \texttt{Matplotlib} \cite{hunter2007matplotlib} and \texttt{seaborn} \cite{waskom2021seaborn}. Machine learning workflows and statistical modeling were implemented using \texttt{scikit-learn} \cite{pedregosa2011scikit}. Finally, this research utilized \texttt{Astropy}, a community-developed core Python package for Astronomy \cite{astropy:2013, astropy:2018, astropy:2022}. The authors acknowledge the vital role of the broader open-source scientific software community in enabling this work.

\section*{Data Availability}
The data that support the findings of this study are available from the corresponding author, IK, upon reasonable request.
 


\bibliographystyle{mnras}
\bibliography{ref} 

@article{kohonen1990self,
  title={The self-organizing map},
  author={Kohonen, Teuvo},
  journal={Proceedings of the IEEE},
  volume={78},
  number={9},
  pages={1464--1480},
  year={1990},
  publisher={IEEE}
}

@inproceedings{villa2017stochastic,
  title={Stochastic self-organizing map variants with the R package SOMbrero},
  author={Villa-Vialaneix, Nathalie},
  booktitle={2017 12th International Workshop on Self-Organizing Maps and Learning Vector Quantization, Clustering and Data Visualization (WSOM)},
  pages={1--7},
  year={2017},
  organization={IEEE}
}

@article{lotz2004new,
  title={A new nonparametric approach to galaxy morphological classification},
  author={Lotz, Jennifer M and Primack, Joel and Madau, Piero},
  journal={The Astronomical Journal},
  volume={128},
  number={1},
  pages={163},
  year={2004},
  publisher={IOP Publishing}
}

@article{rosa2018gradient,
  title={Gradient pattern analysis applied to galaxy morphology},
  author={Rosa, RR and De Carvalho, RR and Sautter, RA and Barchi, PH and Stalder, DH and Moura, TC and Rembold, SB and Morell, DRF and Ferreira, NC},
  journal={Monthly Notices of the Royal Astronomical Society: Letters},
  volume={477},
  number={1},
  pages={L101--L105},
  year={2018},
  publisher={Oxford University Press}
}

@inproceedings{abadi2016tensorflow,
  title={$\{$TensorFlow$\}$: a system for $\{$Large-Scale$\}$ machine learning},
  author={Abadi, Mart{\'\i}n and Barham, Paul and Chen, Jianmin and Chen, Zhifeng and Davis, Andy and Dean, Jeffrey and Devin, Matthieu and Ghemawat, Sanjay and Irving, Geoffrey and Isard, Michael and others},
  booktitle={12th USENIX symposium on operating systems design and implementation (OSDI 16)},
  pages={265--283},
  year={2016}
}

@misc{bishop2007pattern,
  title={Pattern recognition and machine learning (information science and statistics)},
  author={Bishop, Christopher M},
  year={2007},
  publisher={Springer New York, NY, USA}
}

@article{coe2012clash,
  title={Clash: precise new constraints on the mass profile of the galaxy cluster a2261},
  author={Coe, Dan and Umetsu, Keiichi and Zitrin, Adi and Donahue, Megan and Medezinski, Elinor and Postman, Marc and Carrasco, Mauricio and Anguita, Timo and Geller, Margaret J and Rines, Kenneth J and others},
  journal={The Astrophysical Journal},
  volume={757},
  number={1},
  pages={22},
  year={2012},
  publisher={IOP Publishing}
}

@article{abraham2003,
  title={The Gini coefficient as a morphological measurement of galaxies},
  author={Abraham, Roberto G. and others},
  journal={The Astrophysical Journal},
  volume={588},
  number={2},
  pages={218},
  year={2003},
  publisher={IOP Publishing}
}

@article{kohonen1982analysis,
  title={Analysis of a simple self-organizing process},
  author={Kohonen, Teuvo},
  journal={Biological cybernetics},
  volume={44},
  number={2},
  pages={135--140},
  year={1982},
  publisher={Springer}
}

@article{kohonen1982self,
  title={Self-organized formation of topologically correct feature maps},
  author={Kohonen, Teuvo},
  journal={Biological cybernetics},
  volume={43},
  number={1},
  pages={59--69},
  year={1982},
  publisher={Springer}
}

@article{scikit-learn,
 title={Scikit-learn: Machine Learning in {P}ython},
 author={Pedregosa, F. and Varoquaux, G. and Gramfort, A. and Michel, V.
         and Thirion, B. and Grisel, O. and Blondel, M. and Prettenhofer, P.
         and Weiss, R. and Dubourg, V. and Vanderplas, J. and Passos, A. and
         Cournapeau, D. and Brucher, M. and Perrot, M. and Duchesnay, E.},
 journal={Journal of Machine Learning Research},
 volume={12},
 pages={2825--2830},
 year={2011}
}

@article{vialaneix2022package,
  title={Package ‘SOMbrero’},
  author={Vialaneix, Nathalie and Maigne, Elise and Mariette, Jerome and Olteanu, Madalina and Rossi, Fabrice},
  year={2022},
  journal={CRAN}
}

@ARTICLE{kolesnikov_paper1,
       author = {{Kolesnikov}, I. and {Sampaio}, V.~M. and {de Carvalho}, R.~R. and {Conselice}, C. and {Rembold}, S.~B. and {Mendes}, C.~L. and {Rosa}, R.~R.},
        title = "{Unveiling galaxy morphology through an unsupervised-supervised hybrid approach}",
      journal = {\mnras},
         year = 2024,
        month = feb,
       volume = {528},
       number = {1},
        pages = {82-107},
          doi = {10.1093/mnras/stad3934},
archivePrefix = {arXiv},
       eprint = {2401.08906},
 primaryClass = {astro-ph.IM},
       adsurl = {https://ui.adsabs.harvard.edu/abs/2024MNRAS.528...82K}
}

@article{kolesnikov_paper2,
    author = {Kolesnikov, I and Sampaio, V M and de Carvalho, R R and Conselice, C},
    title = {Galaxy morphology in CANDELS: addressing evolutionary changes across 0.2 ⩽ z ⩽ 2.4 with hybrid classification approach},
    journal = {Monthly Notices of the Royal Astronomical Society},
    volume = {539},
    number = {3},
    pages = {2765-2779},
    year = {2025},
    month = {04},
    issn = {0035-8711},
    doi = {10.1093/mnras/staf625},
    url = {https://doi.org/10.1093/mnras/staf625},
    eprint = {https://academic.oup.com/mnras/article-pdf/539/3/2765/62941619/staf625.pdf},
}

@article{White1978,
  author  = {White, S. D. M. and Rees, M. J.},
  title   = {Core condensation in heavy halos: a two-stage theory for galaxy formation and clustering},
  journal = {MNRAS},
  year    = {1978},
  volume  = {183},
  pages   = {341--358}
}

@article{Blumenthal1984,
  author  = {Blumenthal, G. R. and Faber, S. M. and Primack, J. R. and Rees, M. J.},
  title   = {Formation of galaxies and large-scale structure with cold dark matter},
  journal = {Nature},
  year    = {1984},
  volume  = {311},
  pages   = {517--525}
}

@article{Davis1985,
  author  = {Davis, M. and Efstathiou, G. and Frenk, C. S. and White, S. D. M.},
  title   = {The evolution of large-scale structure in a universe dominated by cold dark matter},
  journal = {ApJ},
  year    = {1985},
  volume  = {292},
  pages   = {371--394}
}

@article{Mo1998,
  author  = {Mo, H. J. and Mao, S. and White, S. D. M.},
  title   = {The formation of galactic disks},
  journal = {MNRAS},
  year    = {1998},
  volume  = {295},
  pages   = {319--336}
}

@article{Conselice2003,
  author  = {Conselice, C. J.},
  title   = {The Relationship between Stellar Light Distributions of Galaxies and Their Formation Histories},
  journal = {ApJS},
  year    = {2003},
  volume  = {147},
  pages   = {1--28}
}

@article{Lotz2004,
  author  = {Lotz, J. M. and Primack, J. and Madau, P.},
  title   = {A New Nonparametric Approach to Galaxy Morphological Classification},
  journal = {AJ},
  year    = {2004},
  volume  = {128},
  pages   = {163--182}
}

@article{Conselice2014,
  author  = {Conselice, C. J.},
  title   = {The evolution of galaxy structure over cosmic time},
  journal = {ARA\&A},
  year    = {2014},
  volume  = {52},
  pages   = {291--337}
}

@article{Kartaltepe2015,
  author  = {Kartaltepe, J. S. and others},
  title   = {CANDELS Visual Classifications},
  journal = {ApJS},
  year    = {2015},
  volume  = {221},
  pages   = {11}
}

@article{HuertasCompany2015,
  author  = {Huertas-Company, M. and others},
  title   = {A catalog of visual-like morphologies in the 5 CANDELS fields},
  journal = {ApJS},
  year    = {2015},
  volume  = {221},
  pages   = {8}
}

@article{Bruce2012,
  author  = {Bruce, V. A. and others},
  title   = {The structure of massive galaxies at $1<z<3$},
  journal = {MNRAS},
  year    = {2012},
  volume  = {427},
  pages   = {1666--1692}
}

@article{vanderWel2014,
  author  = {van der Wel, A. and others},
  title   = {3D-HST+CANDELS: The evolution of galaxy structure},
  journal = {ApJ},
  year    = {2014},
  volume  = {788},
  pages   = {28}
}

@article{Robertson2023,
  author  = {Robertson, B. and others},
  title   = {Early galaxy assembly in JWST deep fields},
  journal = {ApJ},
  year    = {2023},
  volume  = {942},
  pages   = {L42}
}

@article{Tohill2024,
  author  = {Tohill, Cl{\'a}r-Br{\'i}d and Bamford, Steven P. and Conselice, Christopher J. and Ferreira, Leonardo and Harvey, Thomas and Adams, Nathan and Austin, Duncan},
  title   = {A Robust Study of High-redshift Galaxies: Unsupervised Machine Learning for Characterizing Morphology with JWST up to z $\sim$ 8},
  journal = {ApJ},
  year    = {2024},
  volume  = {962},
  number  = {2},
  pages   = {164},
  doi     = {10.3847/1538-4357/ad17b8}
}

@article{Dai2023,
  author  = {Dai, Yao and Xu, Jun and Song, Jie and Fang, Guanwen and Zhou, Chichun and Ba, Shuo and Gu, Yizhou and Lin, Zesen and Kong, Xu},
  title   = {The Classification of Galaxy Morphology in the H-band of the COSMOS-DASH Field: A Combination-based Machine-learning Clustering Model},
  journal = {ApJS},
  year    = {2023},
  volume  = {268},
  number  = {2},
  pages   = {34},
  doi     = {10.3847/1538-4365/ace69e}
}

@article{Leclercq2017,
  author  = {Leclercq, Floriane and Bacon, Roland and Wisotzki, Lutz and Mitchell, Peter and Garel, Thibault and Verhamme, Anne and Blaizot, J{\'e}r{\'e}my and Hashimoto, Takuya and Herenz, Edmund Christian and Conseil, Simon and Cantalupo, Sebastiano and Inami, Hanae and Contini, Thierry and Richard, Johan and Maseda, Michael and Schaye, Joop and Marino, Raffaella Anna and Akhlaghi, Mohammad and Brinchmann, Jarle and Carollo, Marcella},
  title   = {The MUSE Hubble Ultra Deep Field Survey. VIII. Extended Ly$\alpha$ haloes around high-z star-forming galaxies},
  journal = {Astronomy \& Astrophysics},
  year    = {2017},
  volume  = {608},
  pages   = {A8},
  doi     = {10.1051/0004-6361/201731480}
}

@article{Wisotzki2018,
  author  = {Wisotzki, L. and Bacon, R. and Brinchmann, J. and Cantalupo, S. and Richter, P. and Schaye, J. and Schmidt, K. B. and Urrutia, T. and Weilbacher, P. M. and Akhlaghi, M. and Bouch{\'e}, N. and Contini, T. and Guiderdoni, B. and Herenz, E. C. and Inami, H. and Kerutt, J. and Leclercq, F. and Marino, R. A. and Maseda, M. and Monreal-Ibero, A. and Nanayakkara, T. and Richard, J. and Saust, R. and Steinmetz, M. and Wendt, M.},
  title   = {Nearly all the sky is covered by Lyman-$\alpha$ emission around high-redshift galaxies},
  journal = {Nature},
  year    = {2018},
  volume  = {562},
  number  = {7726},
  pages   = {229--232},
  doi     = {10.1038/s41586-018-0564-6}
}

@article{Leclercq2020,
  author  = {Leclercq, Floriane and Bacon, Roland and Verhamme, Anne and Garel, Thibault and Blaizot, J{\'e}r{\'e}my and Brinchmann, Jarle and Cantalupo, Sebastiano and Claeyssens, Ad{\'e}la{\"i}de and Conseil, Simon and Contini, Thierry and Hashimoto, Takuya and Herenz, Edmund Christian and Kusakabe, Haruka and Marino, Raffaella Anna and Maseda, Michael and Matthee, Jorryt and Mitchell, Peter and Pezzuli, Gabriele and Richard, Johan and Schmidt, Kasper Borello and Wisotzki, Lutz},
  title   = {The MUSE Hubble Ultra Deep Field Survey. XIII. Spatially resolved spectral properties of Lyman alpha haloes around star-forming galaxies at z > 3},
  journal = {Astronomy \& Astrophysics},
  year    = {2020},
  volume  = {635},
  pages   = {A82},
  doi     = {10.1051/0004-6361/201937339}
}

@article{Vitte2025,
  author  = {Vitte, Elo{\"i}se and Verhamme, Anne and Hibon, Pascale and Leclercq, Floriane and Alcalde Pampliega, Bel{\'e}n and Kerutt, Josephine and Kusakabe, Haruka and Matthee, Jorryt and Guo, Yucheng and Bacon, Roland and Maseda, Michael and Richard, Johan and Pharo, John and Schaye, Joop and Boogaard, Leindert and Nanayakkara, Themiya and Contini, Thierry},
  title   = {The MUSE eXtremely Deep Field: Classifying the spectral shapes of Ly$\alpha$-emitting galaxies},
  journal = {Astronomy \& Astrophysics},
  year    = {2025},
  volume  = {694},
  pages   = {A100},
  doi     = {10.1051/0004-6361/202450426}
}

@article{MadauDickinson2014,
  author  = {Madau, Piero and Dickinson, Mark},
  title   = {Cosmic Star-Formation History},
  journal = {Annual Review of Astronomy and Astrophysics},
  year    = {2014},
  volume  = {52},
  pages   = {415--486},
  doi     = {10.1146/annurev-astro-081811-125615}
}

@article{Bertin1996,
  author = {{Bertin}, E. and {Arnouts}, S.},
  title = "{SExtractor: Software for source extraction.}",
  journal = {Astronomy and Astrophysics Supplement Series},
  year = 1996,
  volume = {117},
  pages = {393}
}

@article{Lee2024,
  author = {{Lee}, Jeong Hwan and {Park}, Changbom and {Hwang}, Ho Seong and {Kwon}, Minseong},
  title = "{Morphology of Galaxies in JWST Fields: Initial Distribution and Evolution of Galaxy Morphology}",
  journal = {The Astrophysical Journal},
  year = 2024,
  volume = {966},
  number = {113},
  pages = {1--24},
  doi = {10.3847/1538-4357/ad3448}
}

@misc{Brammer2023a,
  author = {{Brammer}, G.},
  title = "{grizli (v1.8)}",
  year = 2023,
  publisher = {Zenodo},
  doi = {10.5281/zenodo.7701739}
}

@misc{Brammer2023b,
  author = {{Brammer}, G.},
  title = "{eazy-py (v0.6.7)}",
  year = 2023,
  publisher = {Zenodo},
  doi = {10.5281/zenodo.8268031}
}

@article{Brammer2008,
  author = {{Brammer}, G. B. and {van Dokkum}, P. G. and {Coppi}, P.},
  title = "{EAZY: A Fast, Public Photometric Redshift Code}",
  journal = {The Astrophysical Journal},
  year = 2008,
  volume = {686},
  pages = {1503}
}

@article{Windhorst2022,
  author = {{Windhorst}, R. A. and {Ashcraft}, T. A. and {Broadhurst}, T. J. and others},
  title = "{JWST Proposal, Cycle 1, 2738}",
  journal = {JWST Proposal},
  year = 2022
}

@article{Windhorst2023,
  author = {{Windhorst}, R. A. and {Cohen}, S. H. and {Jansen}, R. A. and others},
  title = "{AJ, 165, 13}",
  journal = {The Astronomical Journal},
  year = 2023,
  volume = {165},
  pages = {13}
}

@article{Finkelstein2021,
  author = {{Finkelstein}, S. L. and {Papovich}, C. and {Pirzkal}, N. and others},
  title = "{JWST Proposal. Cycle 1 2079}",
  journal = {JWST Proposal},
  year = 2021
}

@article{Bagley2024,
  author = {{Bagley}, M. B. and {Pirzkal}, N. and {Finkelstein}, S. L. and others},
  title = "{ApJL, 965, L6}",
  journal = {The Astrophysical Journal Letters},
  year = 2024,
  volume = {965},
  pages = {L6}
}

@article{Finkelstein2017,
  author = {{Finkelstein}, S. L. and {Dickinson}, M. and {Ferguson}, H. C. and others},
  title = "{JWST Proposal, Cycle 0 Early Release Science 1345}",
  journal = {JWST Proposal},
  year = 2017
}

@article{Finkelstein2023,
  author = {{Finkelstein}, S. L. and {Bagley}, M. B. and {Ferguson}, H. C. and others},
  title = "{ApJL, 946, L13}",
  journal = {The Astrophysical Journal Letters},
  year = 2023,
  volume = {946},
  pages = {L13}
}

@article{Dunlop2021,
  author = {{Dunlop}, J. S. and {Abraham}, R. G. and {Ashby}, M. L. N. and others},
  title = "{JWST Proposal, Cycle 1 1837}",
  journal = {JWST Proposal},
  year = 2021
}

@article{Kartaltepe2021,
  author = {{Kartaltepe}, J. and {Casey}, C. M. and {Bagley}, M. and others},
  title = "{JWST Proposal, Cycle 1 1727}",
  journal = {JWST Proposal},
  year = 2021
}

@article{Casey2023,
  author = {{Casey}, C. M. and {Kartaltepe}, J. S. and {Drakos}, N. E. and others},
  title = "{ApJ, 954, 31}",
  journal = {The Astrophysical Journal},
  year = 2023,
  volume = {954},
  pages = {31}
}

@article{Pontoppidan2022,
  author = {{Pontoppidan}, K. M. and {Barrientes}, J. and {Blome}, C. and others},
  title = "{ApJL, 936, L14}",
  journal = {The Astrophysical Journal Letters},
  year = 2022,
  volume = {936},
  pages = {L14}
}

@article{harris2020array,
  title={Array programming with {NumPy}},
  author={Harris, Charles R and Millman, K Jarrod and van der Walt, St{\'e}fan J and Gommers, Ralf and Virtanen, Pauli and Cournapeau, David and Wieser, Eric and Taylor, Julian and Berg, Sebastian and Smith, Nathaniel J and others},
  journal={Nature},
  volume={585},
  number={7825},
  pages={357--362},
  year={2020},
  publisher={Nature Publishing Group}
}

@article{virtanen2020scipy,
  title={{SciPy} 1.0: fundamental algorithms for scientific computing in Python},
  author={Virtanen, Pauli and Gommers, Ralf and Oliphant, Travis E and Haberland, Matt and Reddy, Tyler and Cournapeau, David and Burovski, Evgeni and Peterson, Pearu and Weckesser, Warren and Bright, Jonathan and others},
  journal={Nature Methods},
  volume={17},
  number={3},
  pages={261--272},
  year={2020},
  publisher={Nature Publishing Group}
}

@inproceedings{mckinney2010data,
  title={Data structures for statistical computing in python},
  author={McKinney, Wes and others},
  booktitle={Proceedings of the 9th Python in Science Conference},
  volume={445},
  pages={51--56},
  year={2010},
  organization={Austin, TX}
}

@article{hunter2007matplotlib,
  title={Matplotlib: A 2D graphics environment},
  author={Hunter, John D},
  journal={Computing in Science \& Engineering},
  volume={9},
  number={3},
  pages={90--95},
  year={2007},
  publisher={IEEE}
}

@article{waskom2021seaborn,
  title={seaborn: statistical data visualization},
  author={Waskom, Michael L},
  journal={Journal of Open Source Software},
  volume={6},
  number={60},
  pages={3021},
  year={2021}
}

@article{pedregosa2011scikit,
  title={Scikit-learn: Machine learning in Python},
  author={Pedregosa, Fabian and Varoquaux, Ga{\"e}l and Gramfort, Alexandre and Michel, Vincent and Thirion, Bertrand and Grisel, Olivier and Blondel, Mathieu and Prettenhofer, Peter and Weiss, Ron and Dubourg, Vincent and others},
  journal={Journal of Machine Learning Research},
  volume={12},
  pages={2825--2830},
  year={2011}
}

@article{astropy:2013,
  author={{Astropy Collaboration} and Robitaille, T.~P. and Tollerud, E.~J. and Greenfield, P. and Droettboom, M. and Bray, E. and Aldcroft, T. and Davis, M. and Ginsburg, A. and Price-Whelan, A.~M. and Kerzendorf, W.~E. and Conley, A. and Crighton, N. and Barbary, K. and Muna, D. and Ferguson, H. and Grollier, F. and Parikh, M.~M. and Nair, P.~H. and Unther, H. and Deil, C. and Woillez, J. and Conseil, S. and Kramer, R. and Turner, J.~E.~A. and Singer, L. and Fox, R. and Weaver, B.~A. and Zabalza, V. and Edwards, Z.~I. and Azalee Bostroem, K. and Burke, D.~J. and Craig, M.~W. and Barentsen, G. and Mishra, S. and Rostomac, J. and Tollerud, E.~J. and Greenfield, P. and Droettboom, M. and Bray, E. and Aldcroft, T. and Davis, M. and Ginsburg, A. and Price-Whelan, A.~M. and Kerzendorf, W.~E. and Conley, A. and Crighton, N. and Barbary, K. and Muna, D. and Ferguson, H. and Grollier, F. and Parikh, M.~M. and Nair, P.~H. and Unther, H. and Deil, C. and Woillez, J. and Conseil, S. and Kramer, R. and Turner, J.~E.~A. and Singer, L. and Fox, R. and Weaver, B.~A. and Zabalza, V. and Edwards, Z.~I. and Azalee Bostroem, K. and Burke, D.~J. and Craig, M.~W. and Barentsen, G. and Mishra, S. and Rostomac, J.},
  title={{Astropy: A community Python package for astronomy}},
  journal={Astronomy \& Astrophysics},
  volume={558},
  pages={A33},
  year={2013}
}

@article{astropy:2018,
  author={{Astropy Collaboration} and Price-Whelan, A.~M. and Sip{\H{o}}cz, B.~M. and G{\"u}nther, H.~M. and Lim, P.~L. and Crawford, S.~M. and Conseil, S. and Shupe, D.~L. and Craig, M.~W. and Dencheva, N. and Ginsburg, A. and VanderPlas, J.~T. and Bradley, L.~D. and P{\'e}rez-Su{\'a}rez, D. and de Val-Borro, M. and Aldcroft, T.~L. and Cruz, K.~L. and Robitaille, T.~P. and Tollerud, E.~J. and Ardelean, C. and Babej, T. and Bach, Y.~P. and Barentsen, G. and Barmby, P. and Baumbach, A. and Berry, K.~L. and Biscani, F. and Boquien, M. and Craig, M.~W. and Dencheva, N. and Ginsburg, A. and VanderPlas, J.~T. and Bradley, L.~D. and P{\'e}rez-Su{\'a}rez, D. and de Val-Borro, M. and Aldcroft, T.~L. and Cruz, K.~L. and Robitaille, T.~P. and Tollerud, E.~J.},
  title={{The Astropy Project: Building an Open-science Project and Status of the v2.0 Core Package}},
  journal={The Astronomical Journal},
  volume={156},
  number={3},
  pages={123},
  year={2018}
}

@article{astropy:2022,
  author={{Astropy Collaboration} and Price-Whelan, A.~M. and Lim, P.~L. and Earl, N. and Starkman, N. and Bradley, L. and Shupe, D.~L. and Patil, A.~A. and Corrales, L. and Brasseur, C.~E. and N{\"o}the, M. and Donath, A. and Tollerud, E. and Morris, B.~M. and Ginsburg, A. and Vaher, E. and Weaver, B.~A. and Tocknell, J. and Jamieson, W. and van Kerkwijk, M.~H. and Robitaille, T.~P. and Merry, B. and Bachetti, M. and G{\"u}nther, H.~M. and Aldcroft, T.~L. and Alvarado-Montes, J.~A. and Archibald, A.~M. and B{\'o}di, A. and Bapat, S. and Barentsen, G. and others},
  title={{The Astropy Project: Sustaining and Growing a Community-oriented Open-source Project and the Latest Major Release (v5.0) of the Core Package}},
  journal={The Astrophysical Journal},
  volume={935},
  number={2},
  pages={167},
  year={2022}
}

@ARTICLE{Costantin2023,
       author = {{Costantin}, Luca and {P{\'e}rez-Gonz{\'a}lez}, Pablo G. and {Vega-Ferrero}, Jes{\'u}s and {Huertas-Company}, Marc and {Bisigello}, Laura and {Buitrago}, Fernando and {Bagley}, Micaela B. and {Cleri}, Nikko J. and {Cooper}, Michael C. and {Finkelstein}, Steven L. and {Holwerda}, Benne W. and {Kartaltepe}, Jeyhan S. and {Koekemoer}, Anton M. and {Nelson}, Dylan and {Papovich}, Casey and {Pillepich}, Annalisa and {Pirzkal}, Nor and {Tacchella}, Sandro and {Yung}, L.~Y. Aaron},
        title = "{Expectations of the Size Evolution of Massive Galaxies at 3 {\ensuremath{\leq}} z {\ensuremath{\leq}} 6 from the TNG50 Simulation: The CEERS/JWST View}",
      journal = {\apj},
         year = 2023,
        month = apr,
       volume = {946},
       number = {2},
          eid = {71},
        pages = {71},
          doi = {10.3847/1538-4357/acb926},
archivePrefix = {arXiv},
       eprint = {2208.00007},
 primaryClass = {astro-ph.GA},
       adsurl = {https://ui.adsabs.harvard.edu/abs/2023ApJ...946...71C}
}

@ARTICLE{Nelson2019,
       author = {{Nelson}, Dylan and {Pillepich}, Annalisa and {Springel}, Volker and {Pakmor}, R{\"u}diger and {Weinberger}, Rainer and {Genel}, Shy and {Torrey}, Paul and {Vogelsberger}, Mark and {Marinacci}, Federico and {Hernquist}, Lars},
        title = "{First results from the TNG50 simulation: galactic outflows driven by supernovae and black hole feedback}",
      journal = {\mnras},
         year = 2019,
        month = dec,
       volume = {490},
       number = {3},
        pages = {3234-3261},
          doi = {10.1093/mnras/stz2306},
archivePrefix = {arXiv},
       eprint = {1902.05554},
 primaryClass = {astro-ph.GA},
       adsurl = {https://ui.adsabs.harvard.edu/abs/2019MNRAS.490.3234N}
}

@ARTICLE{Pillepich2019,
       author = {{Pillepich}, Annalisa and {Nelson}, Dylan and {Springel}, Volker and {Pakmor}, R{\"u}diger and {Torrey}, Paul and {Weinberger}, Rainer and {Vogelsberger}, Mark and {Marinacci}, Federico and {Genel}, Shy and {van der Wel}, Arjen and {Hernquist}, Lars},
        title = "{First results from the TNG50 simulation: the evolution of stellar and gaseous discs across cosmic time}",
      journal = {\mnras},
         year = 2019,
        month = dec,
       volume = {490},
       number = {3},
        pages = {3196-3233},
          doi = {10.1093/mnras/stz2338},
archivePrefix = {arXiv},
       eprint = {1902.05553},
 primaryClass = {astro-ph.GA},
       adsurl = {https://ui.adsabs.harvard.edu/abs/2019MNRAS.490.3196P}
}

@article{Verstocken2017,
title = {SKIRT: Hybrid parallelization of radiative transfer simulations},
journal = {Astronomy and Computing},
volume = {20},
pages = {16-33},
year = {2017},
issn = {2213-1337},
doi = {https://doi.org/10.1016/j.ascom.2017.05.003},
url = {https://www.sciencedirect.com/science/article/pii/S221313371730001X},
author = {S. Verstocken and D. {Van De Putte} and P. Camps and M. Baes}
}

@article{Camps2015,
title = {SKIRT: An advanced dust radiative transfer code with a user-friendly architecture},
journal = {Astronomy and Computing},
volume = {9},
pages = {20-33},
year = {2015},
issn = {2213-1337},
doi = {https://doi.org/10.1016/j.ascom.2014.10.004},
url = {https://www.sciencedirect.com/science/article/pii/S2213133714000511},
author = {P. Camps and M. Baes}
}

@article{Camps2020,
title = {SKIRT 9: Redesigning an advanced dust radiative transfer code to allow kinematics, line transfer and polarization by aligned dust grains},
journal = {Astronomy and Computing},
volume = {31},
pages = {100381},
year = {2020},
issn = {2213-1337},
doi = {https://doi.org/10.1016/j.ascom.2020.100381},
url = {https://www.sciencedirect.com/science/article/pii/S2213133720300354},
author = {P. Camps and M. Baes}}

@software{Hilbert2019,
  author       = {Bryan Hilbert and
                  Johannes Sahlmann and
                  Kevin Volk and
                  Shannon Osborne and
                  dthatte and
                  Marshall Perrin and
                  Lauren Chambers and
                  Edward Slavich and
                  Jo Taylor and
                  Erik Tollerud and
                  P. L. Lim},
  title        = {spacetelescope/mirage: First github release},
  month        = oct,
  year         = 2019,
  publisher    = {Zenodo},
  version      = {v1.1.1},
  doi          = {10.5281/zenodo.3519262},
  url          = {https://doi.org/10.5281/zenodo.3519262},
}

@article{Kartaltepe2023,
  author = {Kartaltepe, Jeyhan S. and Rose, Caitlin and Vanderhoof, Brittany N. and McGrath, Elizabeth J. and Costantin, Luca and Cox, Isabella G. and Yung, L. Y. Aaron and Kocevski, Dale D. and Wuyts, Stijn and Ferguson, Henry C. and others},
  title = {CEERS Key Paper. III. The Diversity of Galaxy Structure and Morphology at z = 3--9 with JWST},
  journal = {The Astrophysical Journal Letters},
  year = {2023},
  volume = {946},
  number = {1},
  pages = {L15},
  doi = {10.3847/2041-8213/acad01},
  archivePrefix = {arXiv},
  eprint = {2210.14713},
  primaryClass = {astro-ph.GA}
}

@article{VegaFerrero2024,
  author = {Vega-Ferrero, J. and Huertas-Company, M. and Costantin, L. and P{\'e}rez-Gonz{\'a}lez, P. G. and Sarmiento, R. and Kartaltepe, J. S. and Pillepich, A. and Bagley, M. B. and Finkelstein, S. L. and McGrath, E. J. and Knapen, J. H. and Arrabal Haro, P. and Bell, E. F. and Buitrago, F. and Calabr{\`o}, A. and Dekel, A. and Dickinson, M. and Dom{\'i}nguez S{\'a}nchez, H. and Elbaz, D. and Ferguson, H. C. and Giavalisco, M. and Holwerda, B. W. and Kocevski, D. D. and Koekemoer, A. M. and Pandya, V. and Papovich, C. and Pirzkal, N. and Primack, J. and Yung, L. Y. Aaron},
  title = {On the Nature of Disks at High Redshift Seen by JWST/CEERS with Contrastive Learning and Cosmological Simulations},
  journal = {The Astrophysical Journal},
  year = {2024},
  volume = {961},
  number = {1},
  pages = {51},
  doi = {10.3847/1538-4357/ad05bb},
  archivePrefix = {arXiv},
  eprint = {2302.07277},
  primaryClass = {astro-ph.CO}
}

@ARTICLE{Huertas-Company2024,
       author = {{Huertas-Company}, M. and {Iyer}, K.~G. and {Angeloudi}, E. and et al.},
        title = {Galaxy morphology from z ∼ 6 through the lens of JWST},
      journal = {Astronomy and Astrophysics},
         year = 2024,
       volume = {685},
          eid = {A48},
        pages = {A48},
          doi = {10.1051/0004-6361/202346800}
}

@article{Park2022,
  author = {Park, C. and Lee, J. and Kim, J. and Jeong, D. and Pichon, C. and Gibson, B. K. and Snaith, O. N. and Shin, J. and Kim, Y. and Dubois, Y. and Few, C. G.},
  title = {Formation and Morphology of the First Galaxies in the Cosmic Morning},
  journal = {ApJ},
  volume = {937},
  pages = {15},
  year = {2022},
  doi = {10.3847/1538-4357/ac85b5}
}

@ARTICLE{Sampaio2025,
  author = {{Sampaio}, V.~M. and {Kolesnikov}, I. and {de Carvalho}, R.~R. and {Ferreras}, I. and {Silk}, J.},
  title = {{Morphology across cosmic time: assessing the evolution and interplay of disc- and bulge-dominated galaxies in the CANDELS survey}},
  journal = {Monthly Notices of the Royal Astronomical Society},
  year = {2025},
  month = {aug},
  volume = {541},
  number = {2},
  pages = {1164--1179},
  doi = {10.1093/mnras/staf1050},
  eprint = {2506.12205},
  archivePrefix = {arXiv},
  primaryClass = {astro-ph.GA}
}


\section{Appendix: Metrics overlap in function of redshift}\label{sec:appendix1}
\begin{figure}
    \centering
    \includegraphics[width=\columnwidth]{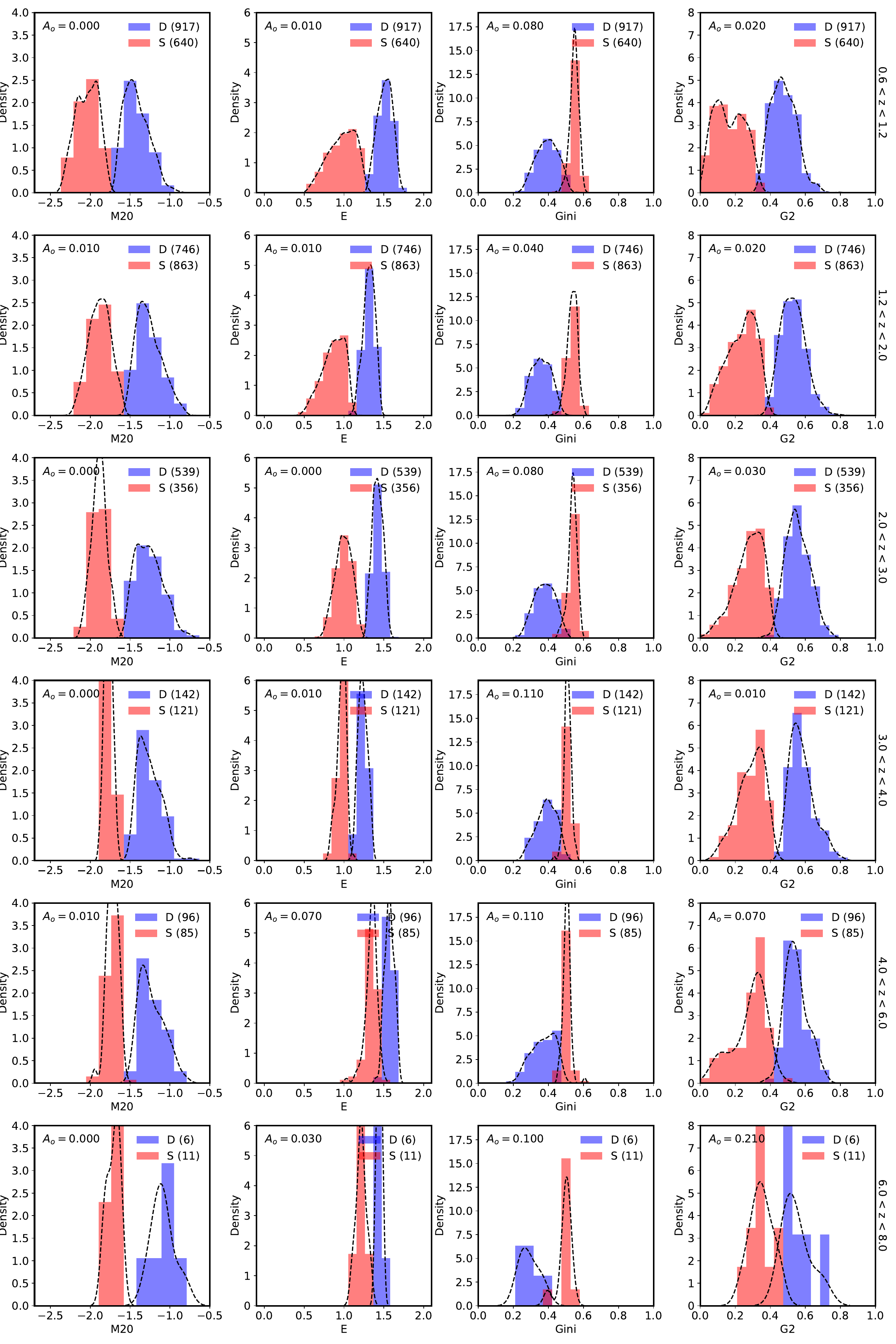}
    \caption{Distributions of the morphological metrics for the prominent clusters as a function of redshift. This visualization should be analyzed in tandem with Figure \ref{fig:all_histograms}.}
    \label{fig:priminents_histograms}
\end{figure}

The Appendix provides additional figures on MEGG system distribution as a  function of redshift, for the prominent cluster galaxies and for the complete dataset. As we can see in Figure \ref{fig:priminents_histograms}, for the prominent cluster galaxies, the metrics exhibit near-perfect class separation across most cases, displaying low overlap area ($\rm A_{\circ}$). Although this stark segregation is an expected consequence of the stringent unsupervised cluster selection criteria, it provides compelling evidence for the robust discriminatory power of non-parametric metrics in morphological classification. 

\begin{figure}
    \centering
    \includegraphics[width=\columnwidth]{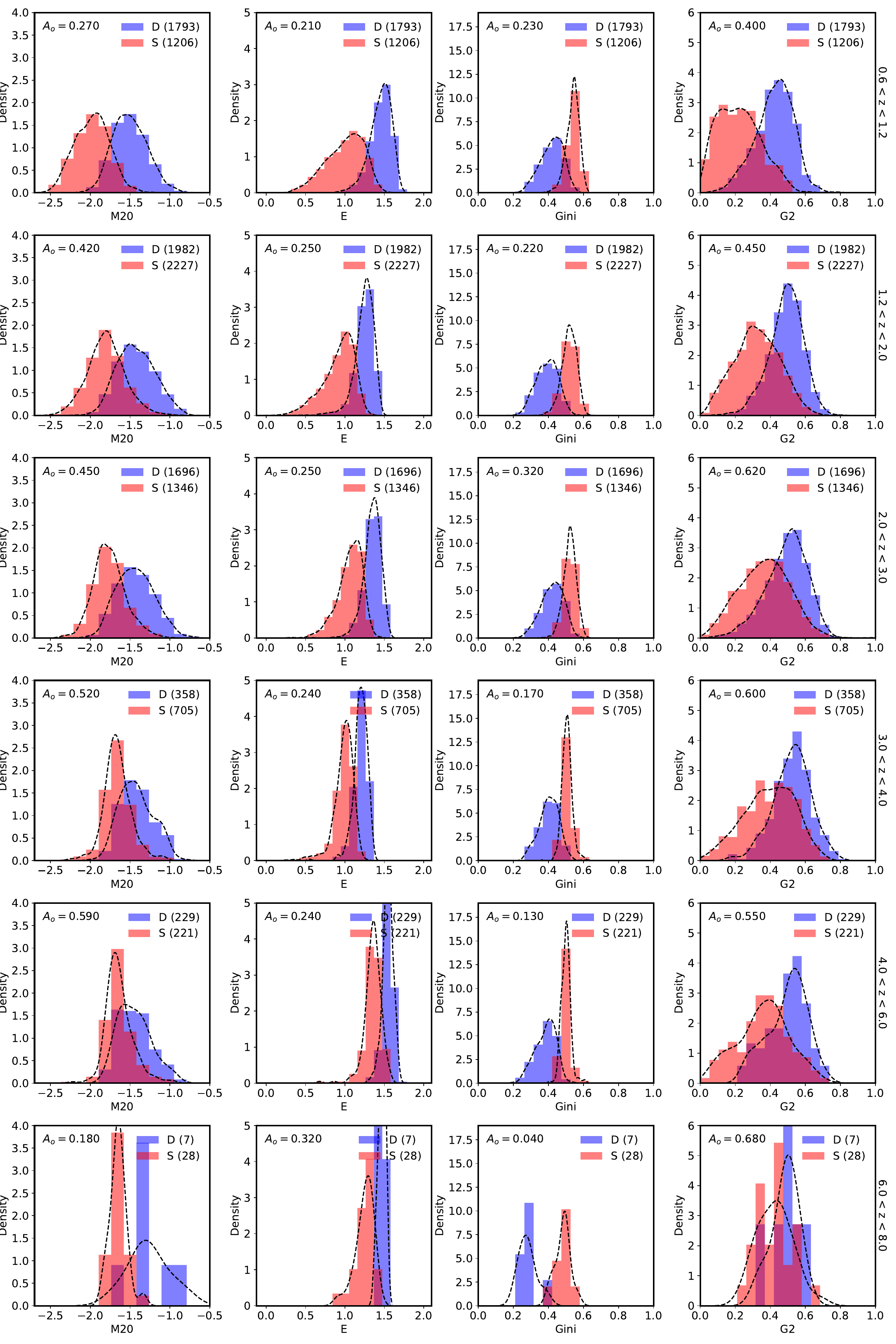}
    \caption{ Distributions of the morphological metrics for the complete galaxy data set as a function of the redshift. }
    \label{fig:all_histograms}
\end{figure}

In Figure \ref{fig:all_histograms}, we show that the parameter-space overlap between the morphological classes, for all galaxies in the sample, is substantially higher across the majority of the redshift bins, when compared to Figure \ref{fig:priminents_histograms}. This increased overlap corroborates our earlier assertion that a comprehensive classification of the entire sample during the initial unsupervised clustering stage is unfeasible. Attempting such a broad classification would inherently introduce significant misclassification noise into the pseudo-labelling process, justifying our strategic decision to isolate and rely exclusively on the prominent clusters for the subsequent supervised training phase.

\bsp	
\label{lastpage}
\end{document}